\documentclass[usenatbib]{mnras}

\usepackage{newtxtext,newtxmath}
\usepackage{verbatim}
\usepackage[T1]{fontenc}
\usepackage{ae,aecompl}
\usepackage{tabularx}

\usepackage{graphicx}	% Including figure files
\usepackage{amsmath}	% Advanced maths commands
\usepackage{xcolor}

\newcommand{\lcdm}{$\Lambda$CDM}
\newcommand{\rvir}{$r_{\rm vir}$}
\newcommand{\mvir}{$M_{\rm vir}$}
\newcommand{\msun}{$M_{\odot}$}
\newcommand{\mstar}{$M_{\star}$}
\newcommand{\atotr}{$a_{\rm tot}(r)$}
\newcommand{\abarr}{$a_{\rm bar}(r)$}
\newcommand{\atot}{$a_{\rm tot}$}
\newcommand{\abar}{$a_{\rm bar}$}

\newcommand{\firetwo}{\texttt{FIRE-2}}
\newcommand{\fbar}{f_{\rm b}}

\title[Hooks and Bends in the RAR]{Hooks \& Bends in the Radial Acceleration Relation: Discriminatory Tests for Dark Matter and MOND}

\author[F. J. Mercado et al.]
{\parbox{17.5cm}{
Francisco J. Mercado$^{1,2,3}$\thanks{E-mail: francisco.mercado@pomona.edu}, James S. Bullock$^{3}$, Jorge Moreno$^{1,4}$, Michael Boylan-Kolchin$^{5}$, Philip F. Hopkins$^{2}$, Andrew Wetzel$^{6}$, Claude-Andr\'e Faucher-Gigu\`ere$^{7}$ and Jenna Samuel$^{5}$}\vspace{0.3cm}\\
$^{1}$Department of Physics and Astronomy, Pomona College, Claremont, CA 91711, USA\\
$^{2}$TAPIR, Mailcode 350-17, California Institute of Technology, Pasadena, CA 91125, USA\\
$^{3}$Department of Physics and Astronomy, University of California Irvine, CA 92697, USA \\
$^{4}$Center for Computational Astrophysics, Flatiron Institute, 162 Fifth Avenue, New York, NY 10010, USA\\
$^{5}$Department of Astronomy, The University of Texas at Austin, 2515 Speedway Stop C1400, Austin, TX 78712, USA\\
$^{6}$Department of Physics, University of California, Davis, CA 95616, USA\\
$^{7}$Department of Physics and Astronomy and CIERA, Northwestern University, 2145 Sheridan Road, Evanston, IL 60208, USA\\
}

\date{Accepted XXX. Received YYY; in original form ZZZ}

\pubyear{2023}

\begin{document}
\label{firstpage}
\pagerange{\pageref{firstpage}--\pageref{lastpage}}
\maketitle

%%%%%%%%%%%%%%%%%%%%%%%%%%%%%%%%%%%%%%%%%%%%%%%%%%%%%%%%%%%%%%%%%%%%%
%%%%%%%%%%%%%%%%%%%%%%%%%%% ABSTRACT %%%%%%%%%%%%%%%%%%%%%%%%%%%%%%%%
%%%%%%%%%%%%%%%%%%%%%%%%%%%%%%%%%%%%%%%%%%%%%%%%%%%%%%%%%%%%%%%%%%%%%

\begin{abstract}
The Radial Acceleration Relation (RAR) connects the total gravitational acceleration of a galaxy at a given radius, \atotr, with that accounted for by baryons at the same radius, \abarr. The shape and tightness of the RAR for rotationally-supported galaxies have characteristics in line with MOdified Newtonian Dynamics (MOND) and can also arise within the Cosmological Constant + Cold Dark Matter (\lcdm) paradigm. We use zoom simulations of 20 galaxies with stellar masses of \mstar\ $\simeq 10^{7-11}$ \msun\ to study the RAR in the \firetwo\ simulations. We highlight the existence of simulated galaxies with non-monotonic RAR tracks that ``hook'' down from the average relation. These hooks are challenging to explain in Modified Inertia theories of MOND, but naturally arise in all of our \lcdm-simulated galaxies that are dark-matter dominated at small radii and have feedback-induced cores in their dark matter haloes. We show, analytically and numerically, that downward hooks are expected in such cored haloes because they have non-monotonic acceleration profiles. We also extend the relation to accelerations below those traced by disc galaxy rotation curves. In this regime, our simulations exhibit ``bends'' off of the MOND-inspired extrapolation of the RAR, which, at large radii, approach  \atot\ $\approx$ \abar$/\fbar$, where $\fbar$ is the cosmic baryon fraction. Future efforts to search for these hooks and bends in real galaxies will provide interesting tests for MOND and \lcdm.
\end{abstract}

% Select between one and six entries from the list of approved keywords.
% Don't make up new ones.
\begin{keywords}
galaxies: formation -- cosmology: theory
\end{keywords}

%%%%%%%%%%%%%%%%%%%%%%%%%%%%%%%%%%%%%%%%%%%%%%%%%%%%%%%%%%%%%%%%%%%%%%%%%%%%%%%%%
%%%%%%%%%%%%%%%%%%%%%%%%%%%%%%%% INTRODUCTION %%%%%%%%%%%%%%%%%%%%%%%%%%%%%%%%%%%
%%%%%%%%%%%%%%%%%%%%%%%%%%%%%%%%%%%%%%%%%%%%%%%%%%%%%%%%%%%%%%%%%%%%%%%%%%%%%%%%%

\section{Introduction}
 The cosmological constant + cold dark matter (\lcdm) model proposes the existence of non-luminous, collisionless (dark) matter that governs galactic dynamics and is essential for structure formation in the Universe \citep[see review;][]{Salucci2019}. An alternative to \lcdm \, for explaining the dynamics of galaxies is MOdified Newtonian Dynamics \citep[MOND;][]{Milgrom1983a,Milgrom1983b,Milgrom1983c}, which changes Newtonian dynamics below a characteristic acceleration scale $a_0 \sim 10^{-10} \, \rm m \, s^{-2}$ in order to explain galaxy rotation curves without the need for dark matter.  Several empirical ``mass-to-light'' scaling relations have been introduced and discussed in the literature within the context of both \lcdm \, and MOND \citep{FJ1976,TF1997,McGaugh2000,McGaugh2015,McGaugh2016}. Of particular interest is the Radial Acceleration Relation \citep[RAR;][]{McGaugh2016}. 
 
In the original RAR paper, \citet{McGaugh2016} showed that galaxies in the Spitzer Photometry and Accurate Rotation Curve (SPARC) database \citep{Lelli2016} scatter tightly around a monotonic relationship between the centripetal acceleration profile, \atot \, (inferred by rotation curves), and the Newtonian acceleration due to the baryonic matter alone, \abar. These quantities can be expressed in terms of the circular velocity at a given radius, $v_{\rm rot}(r)$ and $v_{\rm bar}(r)$, as 

\begin{equation}
    a_{\rm tot}(r) = \frac{v_{\rm rot}^2(r)}{r}, %  = \frac{G M_{\rm tot}(r)}{r^2} , 
    \label{eq:atot}
\end{equation}
and
\begin{equation}
    a_{\rm bar}(r) = \frac{v_{\rm bar}^2(r)}{r}. % = \frac{G M_{\rm bar}(r)}{r^2}.
    \label{eq:abar}
\end{equation}
\citet{McGaugh2016} provide a fit to the empirical RAR with asymptotic behaviour that tracks the MONDian expectation:
\begin{equation}
    a_{\rm tot}(r) = \frac{a_{\rm bar}(r)}{1 - e^{-\sqrt{a_{\rm bar}(r)/a_0}}} \, ,
    \label{eq:rar}
\end{equation}
where $a_{0} = 1.20 \pm 0.26 \times 10^{-10} \, \rm m \ s^{-2}$. For large accelerations, $a_{\rm bar} \gg a_0$, we have $a_{\rm tot} \propto a_{\rm bar}$.  At small accelerations, $a_{\rm bar} \ll a_0$, the relation approaches the low-acceleration MOND prediction $a_{\rm tot} \propto a_{\rm bar}^{1/2}$.  

In the original MOND paper, \citet{Milgrom1983a} used a Modified Inertia formulation, $a \rightarrow \mu(a) a$. In such a theory, the relationship between \atot \, and \abar \, is one-to-one.  Subsequently, \citet{Bekenstein84} introduced Modified Gravity theories, where $F = {\rm m} a$ remains the same but the gravitational field itself is not Newtonian. In principle, Modified Inertia and Modified Gravity could produce different observable predictions. Thus, much work has gone into attempting to discriminate between the two models in MOND \citep[e.g.][]{JS2018,Petersen2020,Eriksen21,Chae22}. For a review of a broader set of modified theories of gravity, see \citet{Shankaranarayanan2022} .

Within a dark-matter framework like \lcdm \,  there is no \textit{single} parameter that defines a characteristic acceleration scale. However, \citet{Kaplinghat02} show that such a scale can emerge as a consequence of dissipative galaxy formation, albeit with a slightly restrictive group of galaxies \cite[for a rebuttal, see][]{Milgrom02}. The observed RAR provides an even stricter test: how does a tight RAR with the observed normalisation {\em and} shape arise within the context of \lcdm? Several studies employ galaxy formation simulations to show that an RAR does arise without fine tuning in \lcdm~\citep{Santos2016,KW2017,Ludlow2017,Tenneti2018,Garaldi2018,Dutton2019}. Though different simulation groups rely on different implementations of star formation and feedback, they all produce fairly tight RARs, albeit with slightly different median trends and scatter to what is observed. \citet{Wheeler2019} argue that the RAR is an algebraic consequence of the Baryonic Tully Fisher Relation (BTFR). \citet{Grudic2020} provide a picture in which a characteristic acceleration scale emerges from stellar feedback physics such that $a_0$ can be expressed using fundamental constants. More recently, \citet{Paranjape2021} present a framework in which the RAR is a result of the interplay between baryonic feedback physics and the distribution of dark matter in galaxies for accelerations $\rm 10^{-12} \, m \, s^{-2}$ $ \lesssim  \, a_{\rm bar} \, \lesssim \, \rm 10^{-10} \, m \, s^{-2}$. 

Several studies employ dark matter halo abundance matching to build semi-empirical models that result in relations with similar normalisation and scatter to the observed RAR \citep{DiCintio2016,Desmond2017,Navarro2017,Li2022}. Notably, \citet{Santos2016} point out that simulated galaxies with profiles that deviate from the Mass Discrepancy-Acceleration Relation, a precursor to the RAR \citep[see][]{McGaugh2014}, also have slowly-rising rotation curves indicative of feedback-induced cores. \citet{Navarro2017} explicitly shows that feedback-induced dark-matter cores may help explain some outlier points in the RAR in galaxies whose rotation curves suggest the presence of such cores. \citet{Ren19} argue that scatter about the average RAR is better explained with self-interacting dark matter (SIDM), pointing out that the central regions of observed galaxies, where SIDM often predicts cores, demonstrate more scatter about the RAR than outer regions (see their Supplemental Material, Figure S1). Conversely, \citet{Li2022} emphasize that low-mass galaxies with cuspy profiles should have upward-bending ``hook" features that deviate from the observed RAR. These authors show that these upward deviations are amplified when considering adiabatic contraction of a Navarro Frank and White halo \citep[NFW,][]{NFW1997} due to baryonic compression and conclude that other effects, such as stellar feedback, would need to be considered to make more accurate predictions. \citet{Chae22b} builds halo models for SPARC galaxies using published fitting functions that encapsulate broad trends between stellar mass, halo mass, and halo density profile shapes from many state-of-the art \lcdm\ hydrodynamic simulations that include feedback.  With these fitting functions as input, \citet{Chae22b} concludes that the \lcdm\ models produce more scatter along the RAR than seen in the SPARC data. While important, this conclusion builds predictions from published correlations between a single galaxy property (stellar mass) and a subset of parameterised halo properties that cannot capture the full complexity of simulation predictions. This motivates further direct work with hydrodynamic simulations of galaxy formation with feedback-regulated star formation.

Finally, some studies attempt to extend predictions for the RAR down to the very low-acceleration regime. \citet{Oman2020} and \citet{Brouwer2021} predict that the RAR at low accelerations should asymptote to a relation set by the cosmic baryon fraction. However, using galaxy-galaxy lensing at large galactocentric radii of MW-like galaxies, \citet{Brouwer2021} suggest that the RAR continues to follow the relation predicted by MOND at the lowest acceleration scales (see \S \ref{sec:ObsBends} for a more detailed discussion of these results). On the other hand, \citet{Buote19} use X-ray observations (out to $\sim$ 100 kpc) of Mrk 1216 (a compact elliptical galaxy), and NGC 6482 (a fossil group) to claim that the cosmic baryon fraction can be recovered at sufficiently large radii when the hot gas content within that radius is included. This radial behaviour of the baryon fraction results in a RAR that bends below that predicted by MOND.

In this work, we compare the RAR for 20 \firetwo\ \lcdm\ zoom simulations against the empirical RAR for real galaxies. \S \ref{sec:analytic_hooks} describes the RAR as an analytic scaling relation and suggests that non-monotonic ``hooks'' should arise naturally in a dark matter framework with feedback-regulated baryonic physics. \S \ref{sec:sims} introduces our simulations and \S \ref{sec:hooks} demonstrates that these reproduce the observed RAR in aggregate, and also include instances with hook features -- which appear as a result of cored dark matter density profiles in the inner regions of low mass galaxies. In \S \ref{sec:bends}, we use our simulations to discuss ``bends'' in the RAR profiles of galaxies that appear at very low accelerations well beyond the regions probed by galaxy rotation curves. These bends are a consequence of total baryonic mass profiles reaching baryonic closure at large radii. In \S \ref{sec:HooksAndBends}, we provide a discussion of how our results can serve as a basis to test models using the RAR. Finally, \S \ref{sec:conclusions} summarises our results.

\label{sec:intro}

%%%%%%%%%%%%%%%%%%%%%%%%%%%%%%%%%%%%%%%%%%%%%%%%%%%%%%%%%%%%%%%%%%%%%%%%%%%%%%%%%%%%%%%%
%%%%%%%%%%%%%%%%%%%%%%%%%%%%%% ANALYTIC EXPECTATIONS %%%%%%%%%%%%%%%%%%%%%%%%%%%%%%%%%%%
%%%%%%%%%%%%%%%%%%%%%%%%%%%%%%%%%%%%%%%%%%%%%%%%%%%%%%%%%%%%%%%%%%%%%%%%%%%%%%%%%%%%%%%%
\section{Analytic Expectations}

\label{sec:analytic_hooks}

\begin{figure*}
	\includegraphics[width=1.8\columnwidth, trim = 0 0 0 0]{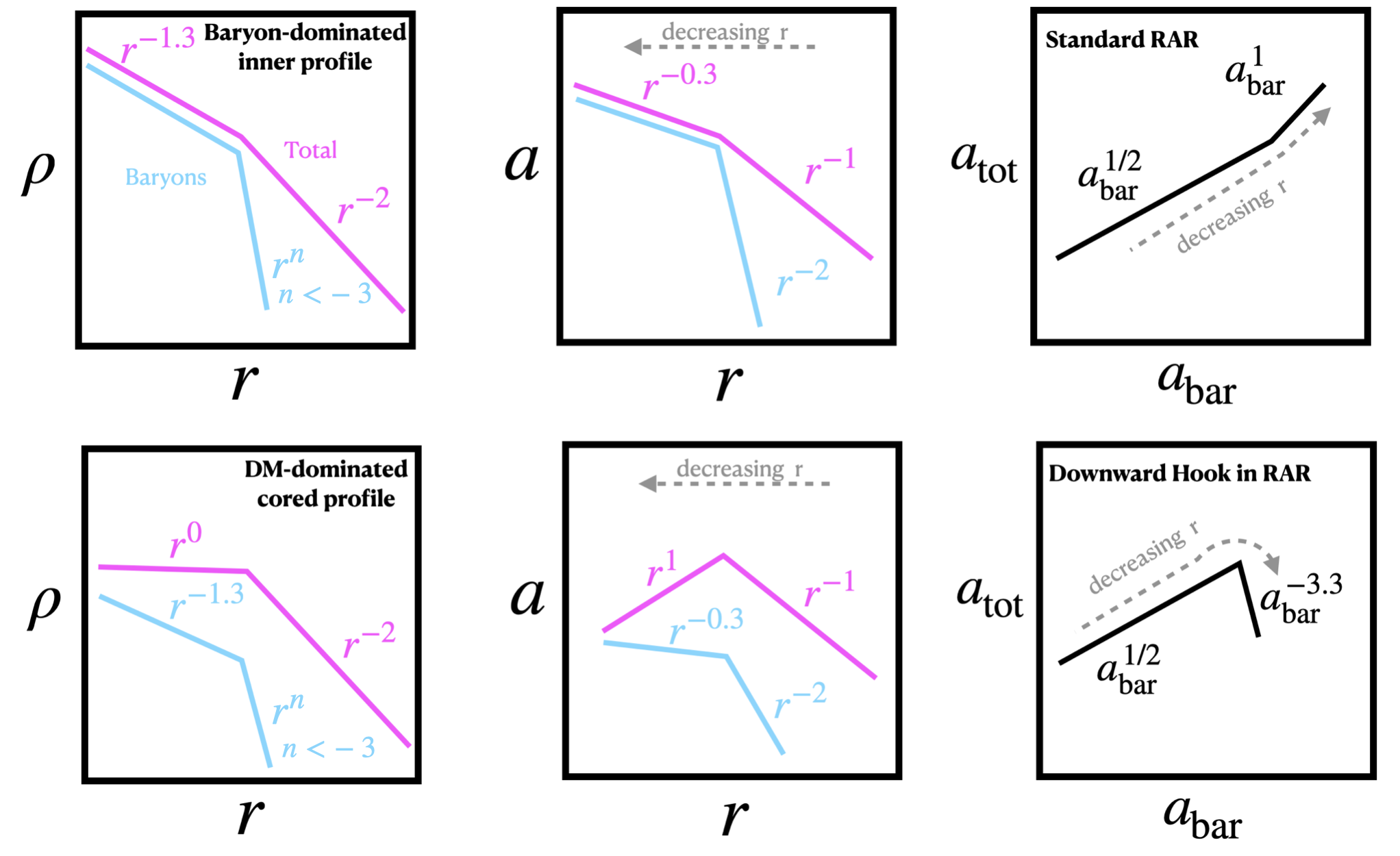}
	\centering
	\caption[]{--- \textbf{\textit{Schematic examples: a standard RAR and a downward hook.}} The upper and lower panels show simple examples of how spherically-symmetric 3D density profiles (left) of total mass distributions (magenta) and baryonic mass distributions (cyan) map to radial acceleration profiles (middle) and ultimately to the RAR (right). Each panel assumes a log-log axis scaling. The dashed grey arrow in the middle and right panels is pointed in the direction of decreasing radius. In the upper panels, we assume a baryon-dominated, inner cuspy profile, and this produces a standard-type RAR relation (upper right).  In the lower set of figures, we assume a dark-matter-dominated inner mass profile, with a cored density distribution.  This assumption gives rise to an RAR profile with a downward hook, of the type shown for real galaxies in Figure \ref{fig:sparc_hooks} and simulated galaxies in Figure \ref{fig:RAR}. See the end of \S \ref{sec:analytic_hooks} for a more detailed description. \textbf{\textit{Takeaway:}} Reasonable assumptions for the density makeup of baryon-dominated galaxies allows us to understand the observed average scaling of the RAR in a natural way (top); these expectations break down for dark-matter dominated galaxies with cored inner dark matter density profiles, which should often deviate from the average scaling (bottom).}
	\label{fig:cartoon}
\end{figure*}
Here we provide a simplified analytic framework to guide expectations. 
Let us assume that the total and baryonic matter exhibit spherically symmetric distributions with cumulative mass profiles $M_{\rm tot}(r)$ and $M_{\rm bar}(r)$, such that \atot $\propto M_{\rm tot}/r^2$ and \abar $\propto M_{\rm bar}/r^2$.  Now, if we characterise the total mass profiles as local power-laws with slopes $p(r)$ that vary slowly with radius:  $M_{\rm tot} \propto r^{p_{\rm tot}}$ and $M_{\rm bar} \propto r^{p_{\rm bar}}$, this allows us to write
\begin{equation}
   a_{\rm tot}  \propto   r^{p_{\rm tot}-2} \: \: {\rm and} \: \:
       a_{\rm bar}  \propto   r^{p_{\rm bar}-2}.
          \label{eq:ascale}
\end{equation}
Note that for radii large enough to contain the total mass, $p(r) \rightarrow 0$, yielding the expected Keplerian scaling $a \propto 1/r^2$ as $r \rightarrow \infty$. Equation \ref{eq:ascale} allows us to write the scaling behaviour of the RAR as
\begin{eqnarray}
   a_{\rm tot}(r) & \propto &   a_{\rm bar}(r)^{m} \: \: ;  \: \:  m \equiv \frac{p_{\rm tot} - 2}{p_{\rm bar} - 2}.
   \label{eq:rarscale}
\end{eqnarray}
For many familiar mass profiles, the acceleration increases monotonically with radius and is always largest at small radii ($p_{\rm bar} < 2$ and $p_{\rm tot} < 2$) with $p(r)$ decreasing as $r$ increases. In such cases, the relationship between \abarr\ and \atotr\ will also be  {\em monotonic}. Note, however, that if the value of $m(r)$ ever changes sign as a function of radius, the relationship between \abarr\ and \atotr\ will no longer be monotonic. The MOND-inspired RAR parameterisation provided by \citet{McGaugh2016} is explicitly monotonic (see our Equation \ref{eq:rar}) and has $m=1$ at large accelerations, $a_{\rm bar} \gg a_0$, and $m=1/2$ at small accelerations, $a_{\rm bar} \ll a_0$.  

We can understand the asymptotic scaling of the RAR for {\em massive} galaxies within a \lcdm\ framework as follows. At small radii and large accelerations, such galaxies are typically baryon dominated \citep{Tollerud2011,Capellari2013,Lovell2018}.  In this regime, $m=1$ occurs naturally because $M_{\rm tot} \simeq M_{\rm bar}$,  $p_{\rm tot} \simeq p_{\rm bar}$,  and $m \simeq 1$.  At large radii and low accelerations, the baryonic acceleration must track the Keplerian expectation with $p_{\rm bar} \simeq 0$. If, as is usually observed, the total rotation curve is flat out to the galaxy's edge, $M_{\rm tot} \propto r$ and  $p_{\rm tot} = 1$.  With $p_{\rm bar} = 0$ and $p_{\rm tot} = 1$ we have $m=1/2$ at large $r$ and small $a$.~\footnote{Whilst this asymptotic behaviour makes sense, it is important to recognise that the observed existence of flat rotation curves below $a_0$ is a key motivation for MOND in the first place.  In the context of \lcdm, the question is whether the flattening occurs as observed.  As discussed in the Introduction and shown in \S \ref{sec:hooks}, many \lcdm\ simulations produce galaxies with acceleration profiles that track the observed RAR from high to low accelerations across the \abar\ $ \simeq \, a_0$ transition remarkably well.}
 
Now consider galaxies that are dark-matter dominated in their centres, as is often the case for low-mass galaxies \citep{Carignan1988,Martimbeau1994,deBlok2997}. In this limit, $M_{\rm bar}(r) \ll M_{\rm tot}(r) \simeq M_{\rm dm}(r)$, where $M_{\rm dm}(r)$ is the dark matter mass distribution. If the dark matter follows a density profile of the form $\rho_{\rm dm} \propto r^{-n}$ at small radii, then in this limit $p_{\rm tot} \simeq  p_{\rm dm} \simeq 3 - n$.   For an NFW-like ``cuspy'' profile \citep{NFW1997} we have $n \rightarrow 1$ at radii smaller than the halo scale radius, which gives $p_{\rm tot} \rightarrow 2$ at small radii.  Interestingly, baryons arrayed in an exponential disc have $p_{\rm bar} \rightarrow 2$ for radii much smaller than the galaxy scale radius.  However, since galaxy scale radii are typically smaller than dark matter scale radii, we expect $p_{\rm bar} \lesssim p_{\rm tot} \simeq 2$ such that $m$ is close to, but less than, unity at the centres of dark-matter-dominated galaxies: $1/2 < m \lesssim 1$.  We refer the reader to \citet{Navarro2017} for a more thorough discussion of how the RAR scaling arises within cuspy dark matter haloes.

Whilst the above discussion may help us to explain {\em on average} why situations with $m \sim 1/2$ at large $r$ (low $a$) and $m \sim 1$ at small $r$ (high $a$) may arise in a \lcdm\ context,  the argument is much less robust for dark-matter dominated galaxies than for baryon-dominated galaxies where $m=1$ is achieved by definition.  Specifically, if at any point along the acceleration profiles of a galaxy, the value of the quantity $m = (p_{\rm tot} - 2)/(p_{\rm bar} -2)$ in Equation \ref{eq:rarscale} changes sign from positive to negative as we approach the inner galaxy, then a ``hook'' in the RAR would emerge. Given that we expect both $p_{\rm bar} \approx 2$ and $p_{\rm tot} \approx 2$ to be reasonable values at small radii in dark-matter-dominated galaxies, it would be surprising if cases {\em never} occurred where one of the slopes had $p \gtrsim 2$ and the other had $p \lesssim 2$ such that hooks appeared.  For example, if we have a dark-matter dominated galaxy where the inner dark matter profile was core-like, with $\rho_{\rm dm} \propto r^{-n}$ and $n <1$, then this will give $p_{\rm tot} >2$ and provide the conditions where a non-monotonic, downward hook would arise. 

Figure \ref{fig:cartoon} displays two schematic examples of how the density distributions (left panels) of baryons (cyan) and total matter (magenta) translate into acceleration profiles (middle panels) and the RAR (right panels). The upper panels correspond to a ``standard RAR'' whilst the lower panels display a ``downward hook''.  In both cases we assume the same large-$r$ behaviour for the baryons: the density falls off quickly with $r$, such that the baryonic acceleration is Keplerian\footnote{The precise slope of the baryonic density profile at large $r$ does not matter as long as it large enough (steeper than $r^{-3}$) to contain the majority of the baryonic mass within a finite radius, which will drive baryonic acceleration towards the Keplerian behaviour beyond that point.} with \abar\ $\propto r^{-2}$.  We also assume that the total density profile produces a flat rotation curve at large radii, with $\rho_{\rm tot} \propto r^{-2}$ and \atot\ $\propto r^{-1}$.  These assumptions produce the familiar low-acceleration behaviour in the RAR: \atot\ $\propto$ \abar$^{1/2}$. 

In the upper panels, labeled ``baryon-dominated inner profile,'' we assume a total density profile dominated by baryons at small radii, with a cuspy inner slope $\rho_{\rm tot} \simeq \rho_{\rm bar} \propto r^{-1.3}$. The specific value of the cusp slope is not important, only that it is steeper than $r^{-1}$, which produces a monotonic acceleration profile.  With this specific choice we obtain $a_{\rm tot} \simeq a_{\rm bar} \propto r^{-0.3}$ and  \atot\ $\propto$ \abar\ at large \abar. In the upper right panel we see that the two asymptotic slopes match the fiducial RAR values.

In the lower panels, labeled ``DM-dominated cored profile,'' we assume that the total density profile is dominated by a cored dark matter halo with $\rho_{\rm tot} \simeq \rho_{\rm dm} \propto r^0$.  This implies that the total acceleration profile obeys \atot\ $\propto r^{1}$ at small $r$, which immediately demands that the \atotr\ profile is non-monotonic with radius.  If we assume that the baryonic profile {\em is} monotonic, following the same scaling assumed in the upper panel, then this produces a non-monotonic RAR with \atot\ $\propto$ \abar$^{-3.3}$ at large \abar\ (corresponding to small radii). The shape this makes in the lower right panel is what we refer to as a downward ``hook.''

The aforementioned case illustrates one specific example of a scenario that gives rise to a hook-like feature in the RAR.  Generally, we expect hooks to occur in the RAR when some combination of \abarr\ and/or \atotr\ are non-monotonic\footnote{In principle, it is possible for both \abarr\ and \atotr\ to be non-monotonic and still produce a monotonic RAR, but that would require extreme fine tuning.}. Whether the hook is upward or downward in the RAR diagram depends on the relative shapes of the acceleration profile.  First, if \atotr\ peaks and \abarr\ does not, then the RAR hook will be {\em downward}: $m \sim 1/2 \rightarrow m<0$ as \abar\ increases.  This is what is illustrated in the bottom panel of Figure \ref{fig:cartoon}.  If \abarr\ peaks and \atotr\ does not, then the hook will be {\em upward}: $m \sim 1/2 \rightarrow m\gg 1$ as \abar\ increases.  If they both peak, we will have {\em downward hooks} if \atotr\ peaks at a larger radius than \abarr.  Conversely, we will have {\em upward hooks} if \abarr\ peaks at a larger radius than \atotr.

%%%%%%%%%%%%%%%%%%%%%%%%%%%%%%%%%%%%%%%%%%%%%%%%%%%%%%%%%%%%%%%%%%%%%%%%%%%%%%
%%%%%%%%%%%%%%%%%%%%%%%%%%%%%% SIMULATIONS %%%%%%%%%%%%%%%%%%%%%%%%%%%%%%%%%%%
%%%%%%%%%%%%%%%%%%%%%%%%%%%%%%%%%%%%%%%%%%%%%%%%%%%%%%%%%%%%%%%%%%%%%%%%%%%%%%

\section{Simulations}

We employ cosmological zoom simulations run with the multi-method gravity plus hydrodynamics code GIZMO \citep{Hopkins15} from the Feedback In Realistic Environment (FIRE\footnote{\href{https://fire.northwestern.edu/}{https://fire.northwestern.edu/}}) project. Our simulations are initialised following the method described in \citet{Onorbe2014} and run using the \firetwo\ feedback implementation \citep{Hopkins18}, utilising the mesh-free Lagrangian Godunov (MFM) method. The MFM approach provides adaptive spatial resolution and maintains conservation of mass, energy, and momentum. The \firetwo\ model includes gas heating and cooling with a temperature range of $T = 10 - 10^{10} \, K$. Gas cooling is due to molecular transitions and metal-line fine structure transitions at low temperatures whilst  cooling at temperatures of $\rm \geq 10^4$ K is due to primordial and metal line cooling and free-free emission. The simulations include a uniform cosmic ionising background \citep{FG2009} and multiple channels of stellar feedback -- including Type II and Type Ia supernovae, winds from OB stars and AGB mass loss, and radiative feedback (photoionisation, photoelectric heating, and radiation pressure). All feedback quantities are tabulated from standard stellar population models {\em without subsequent adjustments or fine-tuning} \citep[][STARBURST99]{Leitherer99}. The simulations generate and track eleven separate chemical species (H, He, C, N, O, Ne, Mg, Si, S, Ca, and Fe) for both gas and stars. Star formation occurs
in self-gravitating, Jeans unstable and self-shielding molecular gas above a threshold density of $n_{\rm crit} \geq 1000$ cm$^{-3}$. After a star particle is formed, it is treated as a single stellar population with a Kroupa Initial Mass Function (IMF) \citep{Kroupa02} with mass and metallicity inherited from its progenitor gas particle. 

\begin{table}
    \centering % used for centering table
        %\begin{tabularx}{\textwidth}{lccccc}
        \begin{tabular}{cccc}
            \hline
            \hline  %inserts double horizontal lines
            $\rm Simulation$  & $\rm Baryonic \, Mass$             & $\rm Virial \, Mass$               & $\rm Virial \, Radius$\\
            $\rm Name$        & [$\rm log(M_{\rm bar}/M_{\odot})$] & [$\rm log(M_{\rm vir}/M_{\odot})$] & [$\rm kpc$]           \\
            %                  & [$\times 10^3 M_{\odot}$] & [$\rm log(M_{\rm bar}/M_{\odot})$] & [$\rm log(M_{\rm vir}/M_{\odot})$] & [$\rm kpc$] \\
      
            \hline 
            Isolated m12's &  &   &  \rule{0pt}{2.6ex} \rule[-0.9ex]{0pt}{0pt}\\
            \hline

            \texttt{m12b}$^{\rm (A)}$    &    11.07    &   12.04   &    335    \rule{0pt}{2.6ex}\\
            \texttt{m12c}$^{\rm (A)}$    &    10.93    &   12.03   &    328     \\
            \texttt{m12f}$^{\rm (B)}$    &    11.10    &   12.10   &    355     \\
            \texttt{m12i}$^{\rm (C)}$    &    10.97    &   11.96   &    314     \\
            \texttt{m12m}$^{\rm (D)}$    &    11.19    &   12.06   &    342     \\
            \texttt{m12w}$^{\rm (E)}$    &    10.85    &   11.92   &    301     \\

            \hline
            Elvis Pairs &  &  & \rule{0pt}{2.6ex} \rule[-0.9ex]{0pt}{0pt}\\
            \hline
            
            \texttt{Romeo}$^{\rm (A)}$    &    11.02   &   12.01   &    317     \rule{0pt}{2.6ex}\\
            \texttt{Juliet}$^{\rm (A)}$   &    10.81   &   11.93   &    302     \\
            \texttt{Thelma}$^{\rm (A)}$   &    11.07   &   12.03   &    332     \\
            \texttt{Louise}$^{\rm (A)}$   &    10.69   &   11.93   &    310     \\
            \texttt{Romulus}$^{\rm (F)}$  &    11.19   &   12.18   &    375     \\
            \texttt{Remus}$^{\rm (F)}$    &    10.87   &   11.99   &    320     \\

            \hline
            m11's &  &  & \rule{0pt}{2.6ex} \rule[-0.9ex]{0pt}{0pt}\\
            \hline

            \texttt{m11d}$^{\rm (G)}$    &    9.81   &   11.42   &    204    \rule{0pt}{2.6ex}\\
            \texttt{m11e}$^{\rm (G)}$    &    9.47   &   11.15   &    166    \\
            \texttt{m11h}$^{\rm (G)}$    &    9.89   &   11.24   &    177    \\
            \texttt{m11i}$^{\rm (G)}$    &    9.37   &   10.83   &    128    \\

            \hline
            m10's &  &  & \rule{0pt}{2.6ex} \rule[-0.9ex]{0pt}{0pt}\\
            \hline
            
            \texttt{m10xb}$^{\rm (H)}$    &    8.46   &   10.35   &    66    \rule{0pt}{2.6ex}\\
            \texttt{m10xc}$^{\rm (H)}$    &    8.75   &   10.50   &    74    \\
            \texttt{m10xd}$^{\rm (H)}$    &    8.19   &   10.59   &    79    \\
            \texttt{m10xe}$^{\rm (H)}$    &    8.97   &   10.66   &    83    \\
            
            \hline
            \hline
            
        \end{tabular}
    \caption{Columns from left to right: (1) Simulations names. The superscript letter corresponds to the reference papers for each simulation. (2) Total baryonic mass within ten percent of the virial radius. (3) Halo virial mass. (4): Halo virial radius. Reference papers for simulations -- A: \citet{GK2019a}; B: \citet{GK2017}; C: \citet{Wetzel2016}; D: \citet{Hopkins18}; E: \citet{Samuel2020}; F: \citet{GK2019b}; G: \citet{Elbadry2018}; H: \citet{Graus2019}. Note that all of our simulations with baryonic masses less than $10^{10}$ \msun\ (the m11's and m10's) have core-like inner dark matter profiles and appear as downward hooks in RAR space.}
    \label{table:sim_table}
\end{table}

In this work, we define dark matter haloes to be spherical systems with viral radii, \rvir, inside which the average density is equal to $\Delta_{\rm vir}(z) \rho_{\rm crit}(z)$. Here, the critical density, $\rho_{\rm crit}$, is defined to be equal to $3H^2(z)/8 \pi G$ and $\Delta_{\rm vir}(z)$ is the redshift-evolving virial overdensity defined by \citet{Bryan1998}, where $\Delta_{\rm vir}(z=0) \sim 100$ in our simulations. The dark matter halo virial mass, \mvir, is then defined as the dark matter mass within \rvir. Finally, we take the stellar mass (\mstar) and the baryonic mass ($M_{\rm bar}$) to be the sum of the stellar mass and baryonic mass within 10 percent of \rvir\, respectively.

Our analysis includes 20 simulated galaxies spanning a stellar mass range of \mstar\ $\sim 10^{7-11}$ \msun\ and a halo virial mass range of \mvir\ $\sim 10^{10-12}$ \msun\ at $z  =  0$. Six galaxies (m12*) are isolated MW-mass analogs and are part of the Latte suite \citep{Wetzel2016,GK2017, Hopkins2017,GK2019a,Samuel2020}. Another six (Romeo \& Juliet, Thelma \& Louise, Romulus \& Remus) are pairs from 3 simulations run as part of the ELVIS on FIRE project \citep{GK2019a,GK2019b}. These galaxies are set in environments with configurations similar to the Local Group (LG)  \citep[just as in][]{GK2014}. Namely, each simulation contains MW and M31 analogues with similar relative separations and velocities to the real MW-M31 pair. The other eight galaxies in our sample (m11* \& m10x*) are isolated and less massive with stellar masses \mstar\ $\simeq \, 10^{7.5-9.6}$ \msun \, and virial masses of \mvir\ $\simeq \, 10^{10.3-11.4}$ \msun \, \citep[see;][]{Elbadry2018,Graus2019}. Table \ref{table:sim_table} indicates properties of our simulated galaxies and relevant references. For the public data release and more information on the core suite of \firetwo\ simulations (m11's \& m12's), please see \citet{Wetzel2023}. Finally, we emphasise in the next section, that all of the m11's and m10's exhibit hook features in RAR space similar to those we discussed in \S \ref{sec:analytic_hooks}.  As discussed in \citet{Lazar2020}, each of these ``hook'' galaxies also has a dark matter profile that is core-like at small radii, with $\rho_{\rm dm} \propto r^n$, $n < 1$.

\label{sec:sims}

%%%%%%%%%%%%%%%%%%%%%%%%%%%%%%%%%%%%%%%%%%%%%%%%%%%%%%%%%%%%%%%%%%%%%%%%%%%%%%%%%%%%%%%%%%%%%%%%%%%%%%%%%%%%%%%%
%%%%%%%%%%%%%%%%%%%%%%%%%%%%%%%%%%%%%%%%%%%%%%%%%% SECTION %%%%%%%%%%%%%%%%%%%%%%%%%%%%%%%%%%%%%%%%%%%%%%%%%%%
%%%%%%%%%%%%%%%%%%%%%%%%%%%%%%%%%%%%%%%%%%%%%%%%%%%%%%%%%%%%%%%%%%%%%%%%%%%%%%%%%%%%%%%%%%%%%%%%%%%%%%%%%%%%%%%%
\section{The Simulated RAR and Hooks at Small Radii}

\begin{figure*}
	\includegraphics[width=1.0\textwidth, trim = 0 0 0 0]{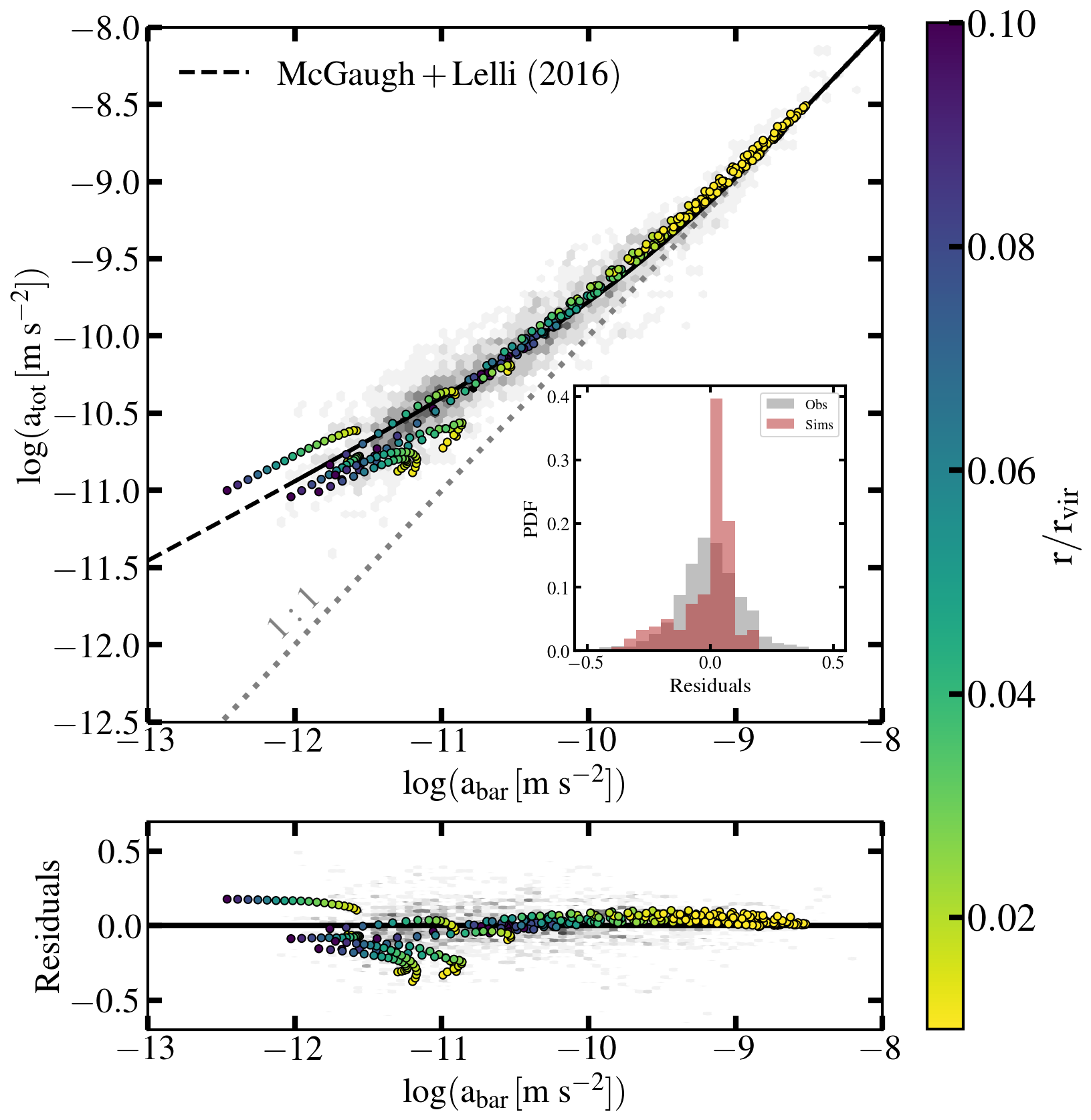}
	\centering
	\caption[]{--- \textbf{\textit{The simulated \& observed radial acceleration relation.}} \textit{Top Panel:} \atot\ versus \abar\ for our simulated galaxy sample (circles) colour-coded by the radius, in units of $r_{\rm vir}$, at which the measurement was performed. The SPARC data used in \citet{McGaugh2016} are illustrated by the grey 2D histogram in the background of this panel. The grey dotted line represents a 1-to-1 relationship whilst  the black line is the fit to the SPARC data introduced by \citet{McGaugh2016}. The dashed portion of the black line represents the same fit extrapolated down to accelerations not probed by the SPARC data. \textit{Inset:} A histogram of the residuals about the black line for the observed and simulated data in grey and red, respectively. \textit{Bottom Panel:} The residuals relative to the \citet{McGaugh2016} fit (black line) as a function of $a_{\rm bar}$ for the simulated and observed data. \textbf{\textit{Takeaway:}} As ensembles, the simulations and observations show strikingly similar RARs.  In addition, several simulated tracks show downward ``hooks,'' reminiscent of the downward hooks highlighted in Figure \ref{fig:sparc_hooks}.}
	\label{fig:RAR}
\end{figure*}

\begin{figure*}
	\includegraphics[width=1.0\textwidth, trim = 0 0 0 0]{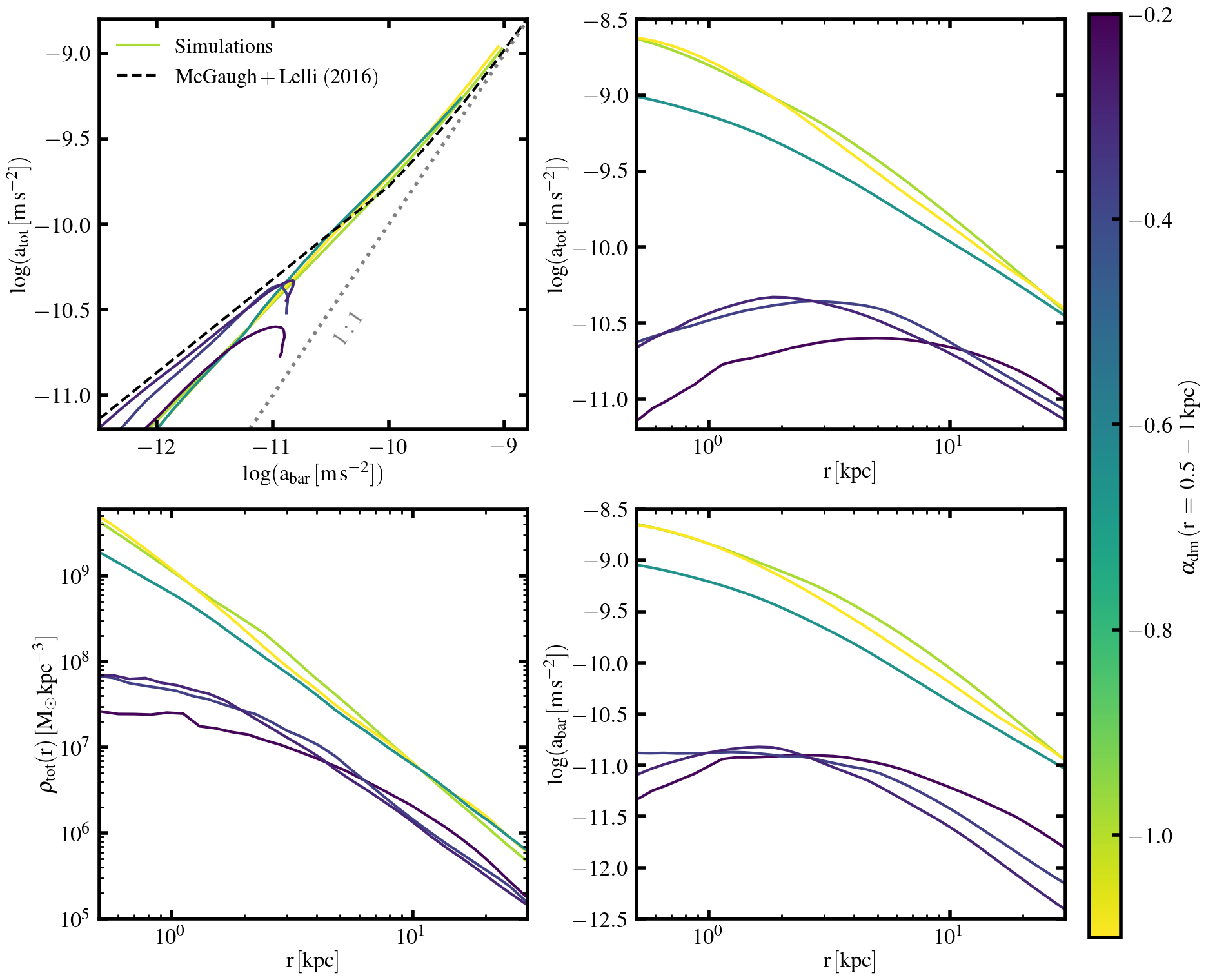}
	\centering
	\caption[]{--- \textbf{\textit{Understanding the hooks.}} The RAR (\textit{top left}), the total radial centripetal acceleration profiles (\textit{top right}), the total radial density profiles (\textit{bottom left}), and the baryonic radial centripetal acceleration (\textit{bottom right}) of a subset of our simulated galaxies. The lines representing the simulated profiles are colour-coded by the log slope of the dark matter profile, $\alpha_{\rm dm} =$ d$\ln \rho_{\rm dm}$/d$\ln r$, measured between 0.5 and 1 kpc, as shown by the colour bar on the right. \textbf{\textit{Takeaway:}} Galaxies with cored central dark matter density profiles also exhibit double-valued total, and sometimes baryonic, radial acceleration profiles and appear as downward hooks in the RAR.}
	\label{fig:profiles}
\end{figure*}

Figure \ref{fig:RAR}, presents a comparison between the simulated and the \textit{observed} RAR \citep{McGaugh2016}. The top panel shows the total centripetal acceleration, \atotr, as a function of the baryonic centripetal acceleration, \abarr, for our simulated sample (circles), colour-coded by the radius of measurement, $r$, in units of $r_{\rm vir}$. For each galaxy we calculated pairs of acceleration values at radii spanning $0.01 \, r_{\rm vir} \, \leq \, r \, \leq \, 0.1 \, r_{\rm vir}$.  We make this choice in order provide a reasonable comparison to the radial rotation curve ranges in the SPARC data reported by \citet{McGaugh2016}. We compute \atotr\ and \abarr\ directly from the simulations using $M_{\rm tot}(r)$ and $M_{\rm bar}(r)$, respectively, assuming spherical symmetry. On the other hand, the same quantities for the SPARC sample (illustrated by the grey 2D histogram in the background of this panel) are inferred by modeling observed galaxy rotation curves and surface brightness profiles. The grey dotted line shows a 1-to-1 relationship whilst  the solid black curve is the MONDian fit (Equation \ref{eq:rar}). The black dashed portion of the curve represents the same fit extrapolated down to accelerations not probed by the SPARC data. The inset shows a histogram of the residuals about the black curve for the observed and simulated data in grey and red, respectively. Finally the bottom panel shows the residuals relative to the \citet{McGaugh2016} fit (black line) as a function of $a_{\rm bar}$ for the simulated and observed data. It is clear that an RAR arises from our simulations that is similar in normalisation to the observed relation. This is in agreement with past work showing that a RAR can arise as a consequence of the \lcdm \, cosmological model \citep{Desmond2017,Ludlow2017,Navarro2017,Dutton2019,Wheeler2019,Grudic2020,Paranjape2021}.

We now draw attention to the hook features in the simulated data near $a_{\rm bar} \, \sim \, 10^{-11} \, \rm m \, s^{-2}$. These features appear well below the characteristic acceleration scale, $a_{0}$, where \atot\ should be proportional to $a_{\rm bar}^m$ with $m \simeq 1/2$ according to MOND. These downward hooks, as predicted in \S \ref{sec:analytic_hooks}, are  manifestly different than the MONDian prediction, and therefore represent an important way to discriminate simulation results like ours against that framework. Figure \ref{fig:profiles} explores the origin of the hook features in our simulations. We plot the RAR (top left), the total radial centripetal acceleration profiles (top right), the total density profiles (bottom left), and the baryonic radial acceleration profiles (bottom right) of a subset of our simulated galaxies (selected to represent the range of simulated galaxy profiles). The lines represent the simulated profiles and are colour-coded according to the log slope of the dark matter profile between 0.5 and 1 kpc. The yellow-to-purple palette represents profile shapes, from cuspy to cored. We conclude that simulated galaxies with cored central dark matter density profiles exhibit double-valued total, and sometimes baryonic, radial acceleration profiles and appear as hooks in the RAR. This is consistent with the analytic expectations discussed in \S \ref{sec:analytic_hooks}.  Ultimately, the predicted hooks in the RAR are a consequence of stellar feedback that redistributes dark matter within the centre-most regions of  low-mass galaxies \citep[see][and references therein]{Ogiya11,Pontzen2012,DiCintio14,Onorbe2015,Chan2015,Lazar2020}. We underscore that, in our simulations, cores arise as a direct consequence of stellar feedback. However, stellar feedback itself was {\it not} introduced to create cores, but rather to regulate star formation.

Finally, we note that non-circular motions of baryons in the inner regions of galaxies can affect their measured rotation curves and their inferred acceleration profiles. For example, {\color{blue} Sands et al. 2024 (in preparation)} use \firetwo\ simulations to show that relaxing the simplifying assumptions that lead to $a \sim GM/r^2$  (including the assumption of spherical symmetry) can yield measured rotation curves that are are inaccurate at the 50\% level or greater in some cases. Thus, providing a more precise comparison to the observations would require inferring velocity profiles from mock-observed images of our simulated galaxies. Given that our goal is to discuss the qualitative nature of the aforementioned hooks in this paper, we leave this for future work.

\label{sec:hooks}

%%%%%%%%%%%%%%%%%%%%%%%%%%%%%%%%%%%%%%%%%%%%%%%%%%%%%%%%%%%%%%%%%%%%%%%%%%%%%%%%%%%%%%%%%%%%%%%%%%%%%%%%%%%%%%%%
%%%%%%%%%%%%%%%%%%%%%%%%%%%%%%%%%%%%%%%%%%%%%%%%%% SECTION %%%%%%%%%%%%%%%%%%%%%%%%%%%%%%%%%%%%%%%%%%%%%%%%%%%
%%%%%%%%%%%%%%%%%%%%%%%%%%%%%%%%%%%%%%%%%%%%%%%%%%%%%%%%%%%%%%%%%%%%%%%%%%%%%%%%%%%%%%%%%%%%%%%%%%%%%%%%%%%%%%%%

\section{Bends at Low Acceleration and Large Radii}

\begin{figure}
	\includegraphics[width=\columnwidth, height=0.28
	\textheight,, trim = 0 0 0 0]{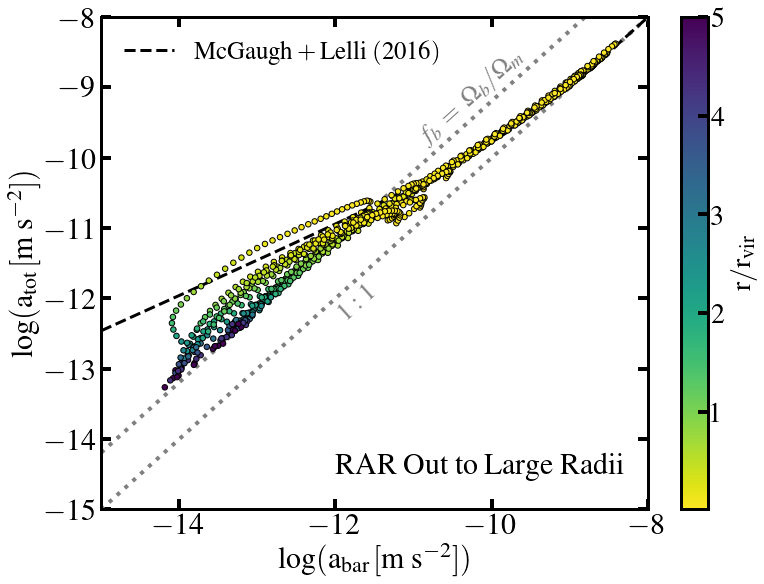}
	\centering
	\caption[]{--- \textbf{\textit{The RAR out to large radii.}} The RAR for our simulated sample (circles) colour-coded by the radius, in units of $r_{\rm vir}$, at which the measurement was performed. The accelerations for each galaxy are provided out to five times the virial radius ($5 \, r_{\rm vir}$). The two dotted grey lines represent a 1-to-1 relationship (labeled as ``1:1'') and a line that tracks the cosmic baryon fraction (labeled ``$f_{b} \, = \, \Omega_{\rm b}/\Omega_{\rm m}$'' ) as \atot $=$ \abar$/\fbar$ with $\fbar = 0.165$. The dashed black line represents the relation provided by \citet{McGaugh2016}. \textbf{\textit{Takeaway:}} The simulated galaxy tracks lie very close to the fit to the SPARC data at accelerations $a_{\rm bary} \, \gtrsim \, 10^{-12} \, \rm m \, s^{-2}$ but bend off at lower accelerations as a result of cosmological homogeneity and the necessity of baryonic closure at large radii.}
	\label{fig:RAR_full}
\end{figure}

\begin{figure}
	\includegraphics[width=\columnwidth, height=0.28
	\textheight,, trim = 0 0 0 0]{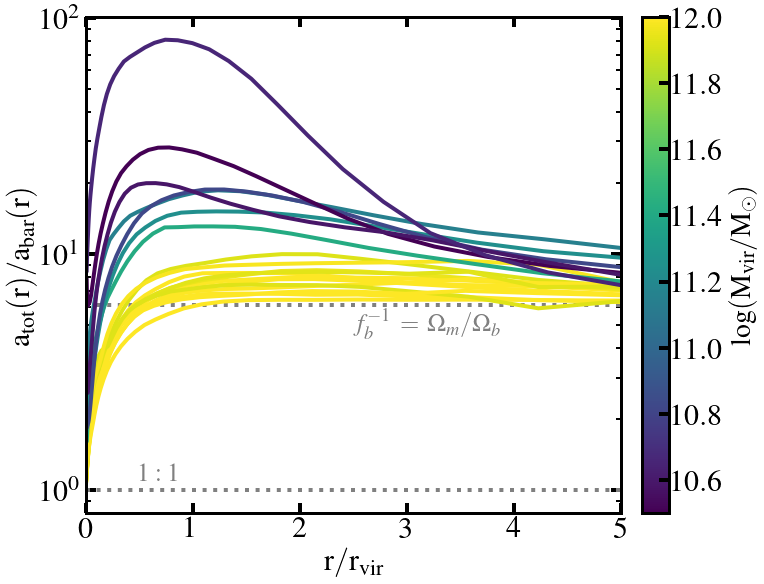}
	\centering
	\caption[]{--- \textbf{\textit{The ratio of \atot\ to \abar\ versus radius.}} The ratio of the total radial acceleration profile to the baryonic radial acceleration profile as a function of radius, normalized to the virial radius ($r_{\rm vir}$) for the simulated galaxies in our sample colour-coded by the virial mass of each galaxy, \mvir. The position on the y-axis where the ratio equals unity and the inverse of the cosmic baryon fraction are represented by horizontal, dotted grey lines. \textbf{\textit{Takeaway:}} Regardless of stellar mass, all galaxies have total to baryonic acceleration ratios that asymptotically approach the inverse baryon fraction at large radius. Galaxies with lower stellar masses have tracks that become baryon deficient at intermediate radii but eventually bend back towards the limit set by cosmology.}
	\label{fig:frac_profile}
\end{figure}

We now extend our analysis out to very large galactocentric radii in order to probe the lowest acceleration scales ($a_{\rm bar} \, \lesssim \, 10^{-12} \, \rm m \, s^{-2}$). Figure \ref{fig:RAR_full} shows the RAR for our simulated sample (circles), colour-coded by the radius, in units of $r_{\rm vir}$, at which we performed our measurements. Here, we provide the accelerations for each galaxy out to five times the virial radius ($5 \, r_{\rm vir}$); the points making up each galaxy track are colour coded by $r/r_{\rm vir}$ as indicated by the colour bar. Note that the baryonic mass here includes stars and \textit{all} gas. This is important because the baryonic mass (and therefore acceleration) at large radii is dominated by diffuse circumgalactic gas \citep[e.g.][]{Li2018,Hafen19}. We note that this diffuse gaseous component is not significantly relevant at the smaller radii traced by galaxy rotation curves, such as those in the SPARC sample and in the regime explored in \S~\ref{sec:hooks}.

The two dotted grey lines represent a 1-to-1 relationship (labelled as ``1:1'') and the line that tracks a 1-to-1 relation with a normalisation set by the cosmic baryon fraction, \atot~$\propto$~\abar$/\fbar$. The dashed black line represents the relation fitted to the SPARC data by \citet{McGaugh2016}, extrapolated down to low accelerations. This figure, similar to Figure \ref{fig:RAR}, shows that the simulated galaxies in our sample follow the fit to the observed RAR fairly well at acceleration scales probed by the SPARC data ($a_{\rm bar} \, \gtrsim \, 10^{-12} \, \rm m \, s^{-2}$). However, at lower accelerations, the simulated galaxies deviate from the extrapolated analytic relation and eventually approach the dotted line set by the cosmic baryon fraction. These ``bends'' are driven by the fact that, at large radii, the fraction of mass in baryons begins to increase towards the cosmic baryon fraction $\fbar$ set by cosmology. By inspecting equations \ref{eq:atot} and \ref{eq:abar}, we eventually reach the limit where $M_{\rm bar} = \fbar \, M_{\rm tot}$, which implies \atot $=$ \abar$/\fbar$. These ``bends'' were previously predicted by \citet{Oman2020} and \citet{Brouwer2021}. Additionally, \citet{Brouwer2021} use galaxy-galaxy lensing at very large galactocentric radii to extend the observed RAR to low accelerations. We include a detailed discussion of the implications of their results in \S \ref{sec:ObsBends}.

In Figure \ref{fig:frac_profile}, we attempt to better understand the bending behaviour by plotting the ratio of the total radial acceleration to the baryonic radial acceleration as a function of galactocentric radius, normalised by the virial radius. The curves are colour-coded by the virial mass. The horizontal dotted grey lines mark the positions on the y-axis where the ratio equals unity (labelled as ``1:1'') and the inverse of the cosmic baryon fraction ($f_{b}^{-1} \, = \, \Omega_{\rm m}/\Omega_{\rm b} \, = \, 6.06$). Notice that all galaxies, regardless of mass, have acceleration profile ratios (or total mass to baryon mass ratios) near unity at small galactocentric radii ($r \ll r_{\rm vir}$), but approach the value set by the cosmic baryon fraction at very large radii ($r \gg r_{\rm vir}$). More massive galaxies (yellow curves) reach baryonic closure by $r \simeq r_{\rm vir}$, whilst their less-massive counterparts (purple curves) have acceleration ratios that stray further from the $f_{b}$ normalisation and only reach baryonic closure at very large radii. This behaviour is driven by the relative power of stellar feedback as a function of galaxy mass. The shallow potential wells of low mass galaxies make it possible for stellar feedback to blow baryons out beyond their virial radii. Additionally, the susceptibility of low mass galaxies to UV background radiation can also prevent the accretion of more baryons. As a result, the baryon fraction lies well below the cosmic value out to quite large radii ($0.5 \, r_{\rm vir} \, < \, r \, < \, 3 \, r_{\rm vir} $) and is not recovered even at $\sim 5\, r_{\rm vir}$ in some cases. On the other hand, the more massive MW-like galaxies have deep enough potential wells that feedback cannot deplete their baryon content as effectively. As a result, the curves of more massive galaxies reach the cosmic baryon fraction scaling at much smaller radii than their less-massive counterparts.

\label{sec:bends}

%%%%%%%%%%%%%%%%%%%%%%%%%%%%%%%%%%%%%%%%%%%%%%%%%%%%%%%%%%%%%%%%%%%%%%%%%%%%%%%%%%%%%%%%%%%%%%%%%%%%%%%%%%%%%%%%
%%%%%%%%%%%%%%%%%%%%%%%%%%%%%%%%%%%%%%%%%%%%%%%%%% SECTION %%%%%%%%%%%%%%%%%%%%%%%%%%%%%%%%%%%%%%%%%%%%%%%%%%%%%
%%%%%%%%%%%%%%%%%%%%%%%%%%%%%%%%%%%%%%%%%%%%%%%%%%%%%%%%%%%%%%%%%%%%%%%%%%%%%%%%%%%%%%%%%%%%%%%%%%%%%%%%%%%%%%%%

\section{Discussion and Implications}
\label{sec:HooksAndBends}

\subsection{Do hooks exist in real galaxies?}
\label{sec:ObsHooks}

\begin{figure*}
	\includegraphics[width=\textwidth, height=0.32
	\textheight,, trim = 0 0 0 0]{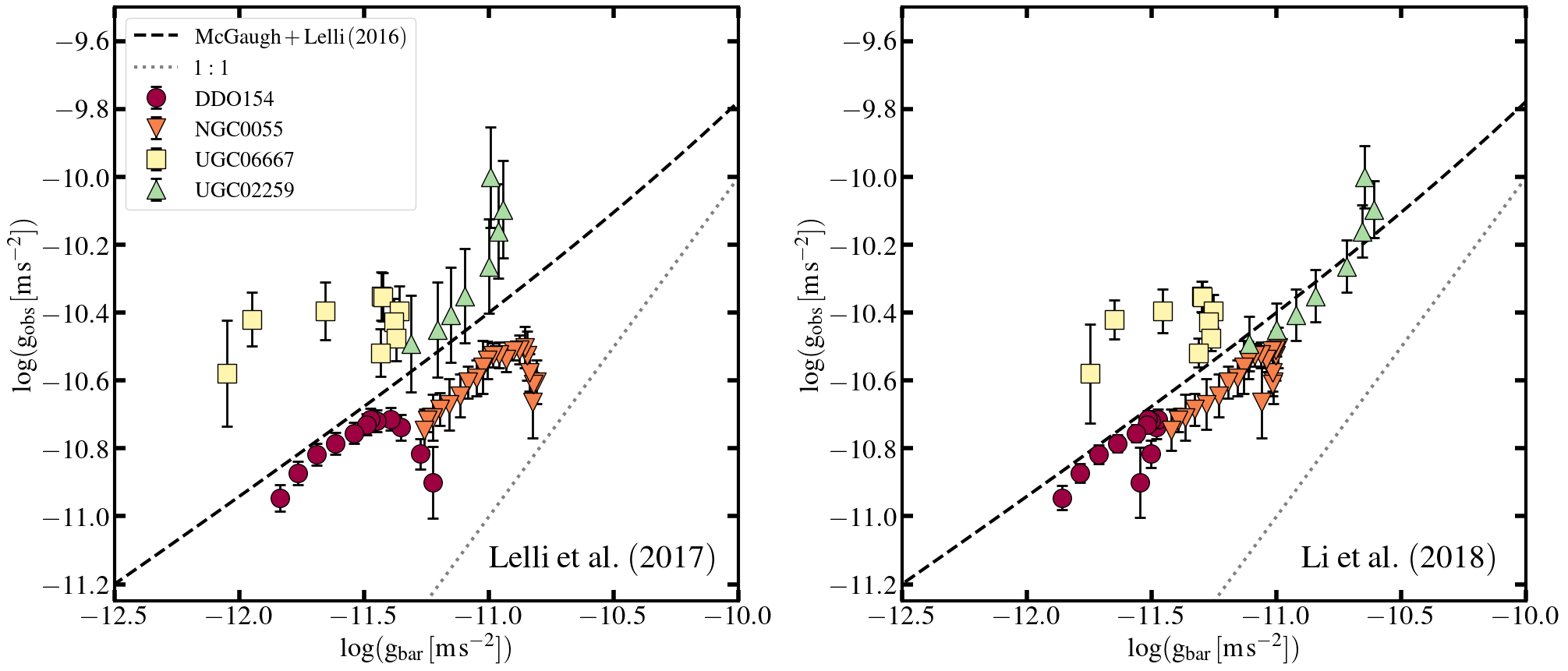}

	\caption[]{--- \textbf{\textit{Potential hooks in real galaxies.}} The red, orange, and yellow sets of points are examples of downward ``hooks''  in the RAR tracks of specific SPARC galaxies; the green points show an example of an upward hook. {\bf Left:} The data on the left come from the \citet{Lelli2017} analysis. {\bf Right:} The data on the right are the same galaxies from the \citet{Li18} analysis, which marginalizes over uncertainties to obtain best-fits to black dashed line. \textbf{\textit{Takeaway:}} We see that after finding best-fit parameters for individual galaxy tracks, the individual RAR tracks of SPARC galaxies lie closer to the MOND-predicted RAR. However, the non-monotonic behaviour at small radii persists.}
	\label{fig:sparc_hooks}
\end{figure*}

\begin{table}
    \centering % used for centering table
    
        %\begin{tabularx}{\textwidth}{lccccc}
        \begin{tabular}{cc|cc}
            \hline
            \hline  %inserts double horizontal lines
            $\rm Galaxy$  &  $\rm Baryonic\, Mass$ & $\rm Galaxy$  &  $\rm Baryonic\, Mass$ \\
            $\rm Name$ & [$\rm log(M_{bar}/M_{\odot})$] & $\rm Name$ & [$\rm log(M_b/M_{\odot})$]\\

            \hline
             Downward Hooks  \rule{0pt}{2.6ex} \rule[-0.9ex]{0pt}{0pt}\\
            \hline

            \texttt{D564-8}          &    7.74     &    \texttt{UGC00731}        &    9.41\\
            \texttt{D631-7}          &    8.68     &    \texttt{UGC04278}        &    9.33\\
            \texttt{DDO154}          &    8.59     &    \texttt{UGC05414}        &    9.12\\
            \texttt{DDO168}          &    8.81     &    \texttt{UGC05764}        &    8.41\\
            \texttt{ESO116-G012}     &    9.55     &    \texttt{UGC05986}        &    9.77\\
            \texttt{F574-1}          &    9.90     &    \texttt{UGC06667}        &    9.25\\
            \texttt{IC2574*}         &    9.28     &    \texttt{UGC06917}        &    9.79\\
            \texttt{KK98-251}        &    8.29     &    \texttt{UGC07089}        &    9.53\\
            \texttt{NGC0055}         &    9.64     &    \texttt{UGC07151}        &    9.29\\
            \texttt{NGC0100}         &    9.63     &    \texttt{UGC07399}        &    9.20 \\
            \texttt{NGC2403}         &    9.97     &    \texttt{UGC07603}        &    8.73\\
            \texttt{NGC3109}         &    8.86     &    \texttt{UGC08837}        &    8.83\\
            \texttt{NGC4010}         &    10.09    &    \texttt{UGCA442}         &    8.62\\

            \hline
             Upward Hooks   \rule{0pt}{2.6ex} \rule[-0.9ex]{0pt}{0pt}\\
            \hline

            \texttt{DDO170}          &    9.10     &    \texttt{NGC4100}         &    10.53\\
            \texttt{NGC0024}         &    9.45     &    \texttt{NGC5585}         &    9.57\\
            \texttt{NGC0247}         &    9.78     &    \texttt{UGC02259}        &    9.18\\
            \texttt{NGC3877}         &    10.58    &    \texttt{UGC04325}        &    9.28\\

            \hline
            \hline
            
        \end{tabular}
    \caption{SPARC Galaxies that we visually identify as having non-monotonic downward hooks in RAR space (top group) and upward hooks in RAR space (bottom group). Examples of these categories are shown as the coloured points in Figure \ref{fig:sparc_hooks}. IC2574, denoted with an asterisk, is the only galaxy in this table whose non-monotonic behaviour does not clearly persist when marginalizing over uncertainties in distance, galaxy inclination and mass-to-light ratio (see \S \ref{sec:ObsHooks}.} Columns 1 \& 3: galaxy names. Columns 2 \& 4: SPARC-quoted baryonic mass.
    \label{table:gal_table}
\end{table}

Given our prediction  that hooks should exist in the high-acceleration (small radius) tracks of dark-matter dominated galaxies with cored inner dark matter profiles, it is interesting to ask if hooks of this kind are observed in real galaxies. Answering this question with confidence is nontrivial because the required information is subject to a number of uncertainties, including disc inclination, galactic distance, and stellar mass-to-light ratios \citep[see][]{Li18,Desmond23}. Nevertheless, we are motivated to share here what is currently known about RAR hooks in the observational realm, which may motivate others to further explore these features in more detail. 

We use two published treatments of SPARC data to visually classify hooks in the individual RAR tracks of galaxies. The first set is a basic presentation published by \citet{Lelli2017}, which assumes fixed mass-to-light ratios. Using this data set, we identify $N=26$ systems ($\sim 15 \%$ of the sample) that display downward hooks and $N=8$ systems ($\sim 5 \%$ of the sample) that display upward hooks. All identified hooks lie outside of the baryon-dominated regime ($M_{\rm bar} \, \lesssim \, 10^{10}$ \msun). Table \ref{table:gal_table} lists these galaxies, along with their baryonic masses.  

The second set of derived acceleration data comes from the analysis published by \citet{Li18}, who marginalise over well-motivated distance, mass-to-light ratio, and inclination angle values to find solutions that best track the monotonic relation given by equation \ref{eq:rar}. After performing a visual inspection of the individual RAR tracks from \citet{Li18}, we find that non-monotonic hooks {\it still exist} for all but one of the galaxies identified in Table \ref{table:gal_table} (IC2574 is listed in our table with an asterisk) -- albeit with tracks that lie closer to the MOND-predicted RAR.

To better-illustrate how these ``observed'' hooks compare between the two analyses, Figure \ref{fig:sparc_hooks} shows RAR tracks for four examples (DDO154, NGC0055, UGC06667 and UGC02259), chosen to represent the diversity of hook behaviour in the SPARC database -- see Figure~\ref{fig:RAR_full} in Appendix~\ref{sec:AllObsHooks} for similar comparisons for the entire set presented in Table~\ref{table:gal_table}. The black dashed line is Equation \ref{eq:rar} and the grey dotted line represents a 1-to-1 relationship. The left panel employs the post-processed data from the first set published \citet{Lelli2017}. The right panel shows the same galaxies with acceleration determinations following the method published by \citep{Li18}, which minimise deviations from equation \ref{eq:rar}. We see that, in all four cases, the hooks persist in the right panel, but lie closer to the observed mean RAR.

\citet{Desmond23} offers improvements upon the \citet{Li18} analysis by simultaneously inferring all nuisance parameters with the properties of the RAR and thus mapping out the degeneracies between all parameters. This analysis reveals an ``underlying'' RAR with a small intrinsic scatter ($\sim 0.034$ dex). Though it is not clear whether this method erases non-monotonic behaviour on a galaxy-by-galaxy basis.

The derived accelerations at small radii can be affected by other factors, such as non-circular motions of baryons at the inner regions of galaxies and finite spatial resolution of the observations. Remarkably, if these hooks are found to be simply artifacts of observational error, this could have important implications for the ``diversity problem'' introduced by \citet{Oman2015}. Specifically, finding that low mass galaxies do not exhibit hook features in their RAR, but rather appear as monotonic tracks that lie along the MOND-predicted RAR, would imply that there is a lack of diversity in their rotation curve shapes. Thus, more detailed work will be needed to determine whether hooks of the kind predicted actually exist at low accelerations and small radii of galaxies. For a detailed discussion of the Mass Discrepancy-Acceleration Relation within the context of the diversity problem, see \citet{Santos2020}.

\subsection{Understanding hooks in a DM context with baryons}
\label{sec:HooksInDM}

Within a DM interpretation, any observed coupling of \atotr\ and \abarr\ outside of the baryon-dominated regime is emergent, driven by galaxy formation processes associated with the collapse of baryons and feedback. This means that the shape of the \atotr\ profile and \abarr\ profile do {\it not} have to track one another. Hooks, in particular, emerge when some combination of \abarr\ and/or \atotr\ are non-monotonic.

If we take the small-radius data for DDO154 and NGC0055 published by \citet{Lelli2017} at face value, they both show non-monotonic \atotr\ profiles accompanied by monotonic \abarr\ profiles. By inspecting Equation \ref{eq:rarscale}, we note that such a situation produces downward hooks, where the value of $m = (p_{\rm tot} - 2)/(p_{\rm bar} -2)$ changes from positive to negative. As the radius decreases, \atotr\ peaks ($p_{\rm tot} = 2$ and $m=0$) and then begins to decline ($p_{\rm tot} >2$, $m<0$), whilst \abarr\ continues to rise ($p_{\rm bar} < 2$). In the same publicly available data set, UGC06667 has double-valued acceleration profiles for both \atotr\ {\em and} \abarr,  but the turnover points occur at different radii. Specifically, \atotr\ peaks and begins to decline at a larger radius than \abarr. This means that as we track the RAR profile from the outer part of UGC06667 inward (from low \abar\ to high \abar), the slope will transition from positive, $m >0$, to negative, $m<0$, as we cross the radius where \atotr\ peaks (where $p_{\rm tot}$ first becomes $> 2$). As can be seen in equation \ref{eq:rarscale}, the slope of the RAR will remain negative ($m<0$) whilst $p_{\rm bar} <2$ and $p_{\rm tot}>2$, until we pass the radius where \abarr\  also peaks (such that now $p_{\rm bar} <2$).  At this point the hook bends back on itself with $m>0$ again.

Finally, UGC02259 (green triangles in Figure~\ref{fig:sparc_hooks}) has a monotonic \atotr\ that  increases as $r$ decreases ($p_{\rm tot} < 2)$, but an  \abarr\ profile that is non-monotonic, peaking at a finite radius where $p_{\rm bar} =2$.  As we follow  \abarr\ from the outside in, it approaches its peak, such that $p_{\rm bar} \rightarrow 2$ (from below) -- whilst $p_{\rm tot} < 2$, which drives $m = (p_{\rm tot} - 2)/(p_{\rm bar} -2) \gg 1$. Such a steep positive slope means that its RAR peels steeply upward away from the average relation before hooking back towards $m <0$ as the \abarr\ profile begins to decline ($p_{\rm bar} > 2$).

Figures \ref{fig:cartoon} and \ref{fig:profiles} illustrate how downward hook features arise in galaxies in our simulations with cored inner dark matter density profiles, which have double-valued total radial acceleration profiles. Cored dark matter profiles arise as a result of star-formation feedback. Note, however, that DM frameworks beyond the standard \lcdm, such as self-interacting dark matter (SIDM), cores can arise even without feedback affecting a galaxy's dark matter radial distribution within a galaxy \citep[e.g.][]{Spergel2000,Vogelsberger12, Rocha13,Kaplinghat16,Tulin18}; and this could provide an alternative way to explain non-monotonic RAR tracks \citep{Ren19}. The detailed shapes of observed tracks could even provide a way to distinguish between feedback-induced CDM cores and SIDM+feedback cores with the same feedback model (Straight et al., in preparation).

We note that \citet{Li2022} use a semianalytic framework to show that upward hook-like features should arise in the RAR tracks of low-mass galaxies if they have cuspy inner profiles. This could happen if star-formation feedback does {\em not} strongly affect the shape of the dark matter halo profile. Detailed simulations like ours do not predict such a situation. However, broadly speaking, \citet{Li2022} make a similar point to ours: the processes that shape the inner structure of galaxies could have important observable imprints on the RAR.

\subsection{What could hooks mean for MOND?}
\label{sec:HooksInMOND}

Non-monotonic relationships between \abar\ and \atot\ are not predicted by Modified Inertia (MI) theories of MOND \citep[e.g.][]{Milgrom1983a,Milgrom22}. \citet{Petersen2020} discuss using the existence of hooks and other features to distinguish between MI and other theories. Interestingly, in Modified Gravity theories \citep[e.g.][]{Bekenstein84,Milgrom2010}, downward hooks {\em can} arise because of the non-spherical symmetry of disk galaxies or external field effects from a galaxy's large-scale cosmic environment \citep[for a thorough investigation\footnote{Whilst \citet{Chae22} predict downward-bending hooks within MOND Modified Gravity models, they do not predict upward hooks.}, see][]{Chae22}. In this context, finding clear evidence for the existence of hooks (or lack thereof) in real galaxies would provide important information for constraining MOND-based theories, in addition to DM-based approaches.

\citet{Eriksen21} discuss how many individual RAR tracks from the SPARC rotation curve database deviate significantly from both Modified Gravity and Modified Inertia predictions, with the existence of both ``cored" and ``cuspy" RAR tracks being difficult for both classes of models to  explain simultaneously.  Of course, the same caveats related to data interpretation discussed in \S 6.1 apply to that analysis. In contrast, \citet{Chae22b} use a statistical sample of SPARC galaxy rotation curves to argue that Modified Gravity with an estimated mean external field correctly predicts the observed statistical relation of accelerations from both the inner and outer parts of rotation curves.

\subsection{Measuring bends in real galaxies}
\label{sec:ObsBends}

In Figure \ref{fig:RAR_full} we point to predicted galaxy tracks that bend off the MOND-inspired fit to the RAR at low accelerations and very large radii, well beyond the radii probed by galaxy rotation curves. This clear departure serves as yet another tool to discriminate between the MOND and \lcdm. \citet{Oman2020} and \citet{Brouwer2021} have previously made the point that the RAR should bend towards \atot\ $\propto$ \abar\ $/f_b$ at large radii for isolated Milky Way size galaxies. \citet{Brouwer2021} extend the RAR to low accelerations by measuring the total acceleration, \atot, using galaxy-galaxy lensing out to very large radii ($\sim 3$ Mpc). If they include only the baryonic mass estimated from the HI gas and stellar mass within $\sim$ 30 kpc to determine \abar, they find that the resulting RAR at large radii continues to follow the relation predicted by MOND (black points in their Figure 4). Importantly, the \citet{Brouwer2021} data include no direct measurement of the hot, ionised gas. They do show that adding an extended ionised gaseous contribution (within $R \sim 100 \, \rm kpc$) to their \abar\ estimates results in a RAR that begins to bend below the expected MONDian relation at very low accelerations (orange points in their Figure 4). These bends are similar in nature to the bends our simulations predict, though not as pronounced. Naively, one would expect circumgalactc gas to extend out to galactocentric radii larger than 100 kpc, which would further affect \abar\ in the regions that \citet{Brouwer2021} probe.

\begin{figure}
	\includegraphics[width=\columnwidth, height=0.28
	\textheight,, trim = 0 0 0 0]{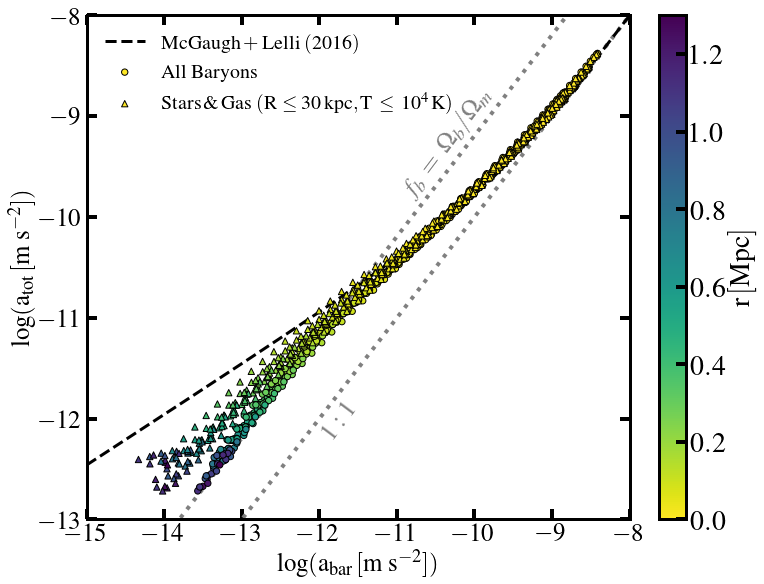}
	\centering
	\caption[]{--- \textbf{\textit{Low acceleration RAR test}} The RAR for our simulated MW and M31 analogs colour-coded by the radius at which the measurement was performed. The accelerations for each galaxy are provided out to 1.3 Mpc ($\sim$ 5 \rvir). The circles represent the same data shown in Figure \ref{fig:RAR_full} while the triangles illustrate how the low-acceleration RAR if we consider only stars and cold gas within 30 kpc of the galactic center to best match the assumptions made in \citet{Brouwer2021}. The two dotted grey lines represent a 1-to-1 relationship (labeled as ``1:1'') and a line that tracks the cosmic baryon fraction (labeled ``$f_{b} \, = \, \Omega_{\rm b}/\Omega_{\rm m}$'' ) as \atot $=$ \abar$/\fbar$ with $\fbar = 0.165$. The dashed black line represents the relation provided by \citet{McGaugh2016}. \textbf{\textit{Takeaway:}} Whilst the points lie closer to the MOND-predicted low-acceleration RAR when considering only baryons at galactocentric radii less than 30 kpc, it is not enough to recover the observed low acceleration RAR. }
	\label{fig:low_a_test}
\end{figure}

Figure \ref{fig:low_a_test} shows the RAR for our simulated MW and M31 analogs down to low accelerations, colour-coded by the radius at which the measurement was made, out to 1.3 Mpc ($\sim$ 5 \rvir). The circles show the RAR tracks for which we consider all baryons in our calculations (the same data are shown in Figure \ref{fig:RAR_full}). The triangles depict the RAR wherein we only include stars and cold gas (T $\leq$ 10$^4$ K) within 30 kpc to calculate \abar\ in order to match the assumptions made in \citet{Brouwer2021}. It is clear that, when we apply such assumptions to our \abar\ calculations, the RAR tracks of our simulated galaxies lie closer to the MONDian prediction. However, it is not enough to recover the observed low-acceleration RAR.

If the galaxies in the sample presented in \citet{Brouwer2021} are truly isolated, then the low-acceleration lensing data is better-described by MOND. However, if the assumption of isolation does not hold (despite the authors' best efforts), determining the true low-acceleration RAR becomes slightly more complicated. On the one hand, the presence of undetected satellite galaxy structure \citep[predicted by Modified Gravity theories;][]{Chae22b} could potentially explain a slight bend seen at low accelerations (orange points in their Figure 4). On the other hand, \citet{Brouwer2021} also state that in the absence of isolation, the combination of lenses and the additional structure along the line-of-site could yield excess surface density profiles, in line with observed density profiles with $\sim r^{-2}$ dependencies at large galactocentric radii -- rather than the intrinsic $\sim r^{-3}$ outer slopes that the lenses have. This alone could make the difference between a low acceleration RAR with a slope of $m\approx$ 1/2 (\atot\ $\propto$ \abar$^{1/2}$), predicted by MOND and a slope of $m \approx$ 1 (\atot\ $\propto$ \abar), predicted by our simulations. Thus, it is important to confirm that an accurate selection of isolated galaxies is employed for such an analysis to be trustworthy.

Focusing on more massive individual galaxies, \citet{Buote19} use X-ray observations to derive RAR relations for compact elliptical galaxy Mrk 1216 and fossil group NGC 6482 out to $\sim 100$ kpc. These authors find that both systems demonstrate offsets and bends away from the RAR when the hot gas contribution is directly measured within that radius. It is important to point out, however, that the implied dark matter halo of Mrk 1216 has a very high concentration ($> 3 \sigma$ outlier), which would require an unusually early formation time in an \lcdm\ context \citep{Dutton2015}.

%%%%%%%%%%%%%%%%%%%%%%%%%%%%%%%%%%%%%%%%%%%%%%%%%%%%%%%%%%%%%%%%%%%%%%%%%%%%%%%%%
%%%%%%%%%%%%%%%%%%%%%%%%%%%%%%%% Summary %%%%%%%%%%%%%%%%%%%%%%%%%%%%%%%%%%%%%
%%%%%%%%%%%%%%%%%%%%%%%%%%%%%%%%%%%%%%%%%%%%%%%%%%%%%%%%%%%%%%%%%%%%%%%%%%%%%%%%%

\section{Summary and Conclusions}

In this paper we examine the radial acceleration relation (\atot\ vs. \abar) tracks of 20 \firetwo\ simulated galaxies. A summary of our results is as follows:

\begin{itemize}
    
    \item When treated as an ensemble, our \firetwo\ galaxies approximately follow follow the empirical RAR at acceleration scales probed by \citet{McGaugh2016}: \atot $\, \gtrsim \, 10 ^{-12} \, \rm m \, s^{-2}$. This supports the idea that the RAR can arise in \lcdm\- based models of galaxy formation (Figure \ref{fig:RAR}).

    \item Downward hook features appear in the RAR tracks of all eight of our simulated galaxies with baryonic masses lower than $10^{10}$ \msun. Each has a cored inner dark matter density profile and the downward hooks are a consequence of them having non-monotonic total radial acceleration profiles (Figure \ref{fig:profiles}). 
    
    \item By extending the RAR to very large radii from galactic centres, our simulations predict relations that bend away from the low-acceleration extrapolation of the \citet{McGaugh2016} fit (Figure \ref{fig:RAR_full}), which is equivalent to the scaling predicted by MOND (equation~\ref{eq:atot}). This behaviour in our simulations is driven by the fact that, at large radii, the total baryonic mass enclosed recovers the cosmic baryon fraction,  $f_{b} \, = \, \Omega_{\rm b}/\Omega_{\rm m} \, = 0.165$ -- ultimately demanding \atot $=$ \abar$/\fbar$ at $r \gg r_{\rm vir}$. This point was first made for by \citet{Oman2020} and \citet{Brouwer2021}.
  
\end{itemize}

Downward hooks (at high acceleration, small radii) and pronounced bends (at low acceleration, large radii) in the RAR tracks of galaxies, as predicted in our \lcdm\ simulations with baryons, are explicitly distinct from the expectations of Modified Inertia theories and Modified Gravity theories, and can thus be used as tests to discriminate between dark matter and MOND. In our simulations, downward hooks are prevalent in low-mass galaxies in the classical regime (with \mstar\ $\simeq 10^{7.5-9.6}$ \msun), which are most prone to feedback-induced core formation \citep[see Figure 13 of][]{Bullock2017}. Such galaxies would be the best targets for follow up studies looking for RAR hooks. 

We identify a number of galaxies in the SPARC database that do appear to display RAR profiles with downward hooks ($\sim 15 \%$) and upward hooks ($\sim 5 \%$) in their individual RAR tracks. This non-monotonic behavior persists even after finding values for galactic distance, inclination and mass-to-light ratios that minimise the deviation about the MOND-predicted RAR \citep[following the methods published by][]{Li18}. Thus, more work is needed to confirm whether or not such hooks do exist at the central regions of real low-mass galaxies.

The best places to look for the outer RAR bends are in the outskirts of high-mass galaxies. Whilst galaxies of all masses in our simulations predict such bends, only around the most massive galaxies do these bends become prominent within the virial radius. Hot gas from X-ray studies and Sunyaev-Zeldovich signals will be easiest to detect around such massive galaxies as their baryonic masses at large radii are dominated by diffuse circumgalactic gas \citep{Li2018,Hafen19}. The existence or absence of bends of this kind at large radii, as discussed by \citet{Brouwer2021}, provide another avenue for testing competing models for the RAR that have been developed to match results at smaller radii.

\label{sec:conclusions}

%%%%%%%%%%%%%%%%%%%%%%%%%%%%%%%%%%%%%%%%%%%%%%%%%%%%%%%%%%%%%%%%%%%%%%%%%%%%%%%%%
%%%%%%%%%%%%%%%%%%%%%%%%%%%%%%%% ACKNOWLEDGMENTS %%%%%%%%%%%%%%%%%%%%%%%%%%%%%%%%
%%%%%%%%%%%%%%%%%%%%%%%%%%%%%%%%%%%%%%%%%%%%%%%%%%%%%%%%%%%%%%%%%%%%%%%%%%%%%%%%%

\section{Acknowledgments}

FJM is funded by the National Science Foundation (NSF) MSP-Ascend Award AST-2316748. FJM and JSB were supported by NSF grant AST-1910965 and NASA grant 80NSSC22K0827. JM is supported by the Hirsch Foundation. CAFG was supported by NSF through grants AST-2108230  and CAREER award AST-1652522; by NASA through grants 17-ATP17-0067 and 21-ATP21-0036; by STScI through grant HST-GO-16730.016-A; and by CXO through grant TM2-23005X. MBK acknowledges support from NSF CAREER award AST-1752913, NSF grants AST-1910346 and AST-2108962, NASA grant 80NSSC22K0827, and HST-AR-15809, HST-GO-15658, HST-GO-15901, HST-GO-15902, HST-AR-16159, HST-GO-16226, HST-GO-16686, HST-AR-17028, and HST-AR-17043 from the Space Telescope Science Institute, which is operated by AURA, Inc., under NASA contract NAS5-26555. Support for PFH was provided by NSF Research Grants 1911233, 20009234, 2108318, NSF CAREER grant 1455342, NASA grants 80NSSC18K0562, HST-AR-15800. JS is supported by the NSF Astronomy and Astrophysics Postdoctoral Fellowship. 

We thank the anonymous referees for their valuable input, which improved the quality of this manuscript. Additionally, we thank Federico Lelli for useful feedback and for providing advice on interacting with SPARC data, Pengfei Li for providing the best-fit parameters used to produce the data in the right hand panel of Figure \ref{fig:sparc_hooks} and Kyle Oman and Isabel Santos-Santos for very useful feedback on the work that we reference within this paper. FJM and JM also thank Andrew Benson and Ana Bonaca for discussions on an earlier draft. The functionalities provided by the following python packages played a critical role in the analysis and visualisation presented in this paper: \textbf{\texttt{matplotlib}} \citep{Hunter2007}, \textbf{\texttt{NumPy}} \citep{vanderWalt2011}, \textbf{\texttt{SciPy}} \citep{Virtanen2020} and \textbf{\texttt{iPython}} \citep{Perez2007}. We honour the invaluable labour of the maintenance and clerical staff at our institutions, whose contributions make our scientific discoveries a reality. This research was conducted on Acjachemen and Tongva Indigenous land.

\section{Data Availability}
The data supporting the plots within this article are available on reasonable request to the corresponding author. A public version of the GIZMO code is available at \href{http://www.tapir.caltech.edu/~phopkins/Site/GIZMO.html}{http://www.tapir.caltech.edu/~phopkins/Site/GIZMO.html}. \firetwo\ simulations are publicly available at \href{http://flathub.flatironinstitute.org/fire}{http://flathub.flatironinstitute.org/fire}. Additional data including simulation snapshots, initial conditions, and derived data products are available at \href{https://fire.northwestern.edu/data/}{https://fire.northwestern.edu/data/}. 

The galaxy sample information and mass models for the 175 SPARC galaxies presented in \citet{Lelli2016} can be found at \href{https://cdsarc.cds.unistra.fr/ftp/J/AJ/152/157/}{https://cdsarc.cds.unistra.fr/ftp/J/AJ/152/157/}. Additionally, a video that shows the individual tracks of each SPARC galaxy references in \citet{Lelli2017} as well as the best fit parameters presented in \citet{Li18} can be found at \href{http://astroweb.cwru.edu/SPARC/}{http://astroweb.cwru.edu/SPARC/}. Finally, we provide the data for all RAR tracks plotted in Figure \ref{fig:AllObsHooks} in the following github repository \href{https://github.com/fjmercado/ObservedRARTracks}{https://github.com/fjmercado/ObservedRARTracks}.

%%%%%%%%%%%%%%%%%%%% REFERENCES %%%%%%%%%%%%%%%%%%

% The best way to enter references is to use BibTeX:

\bibliographystyle{mnras}
\bibliography{paper_refs.bib} 

%%%%%%%%%%%%%%%%%%%%%%%%%%%%%%%%%%%%%%%%%%%%%%%%%%

%%%%%%%%%%%%%%%%% APPENDICES %%%%%%%%%%%%%%%%%%%%%

\appendix
\section{More RAR Track Comparisons}
\label{sec:AllObsHooks}
\begin{figure*}
	\includegraphics[width=1.0\textwidth, trim = 0 0 0 0]{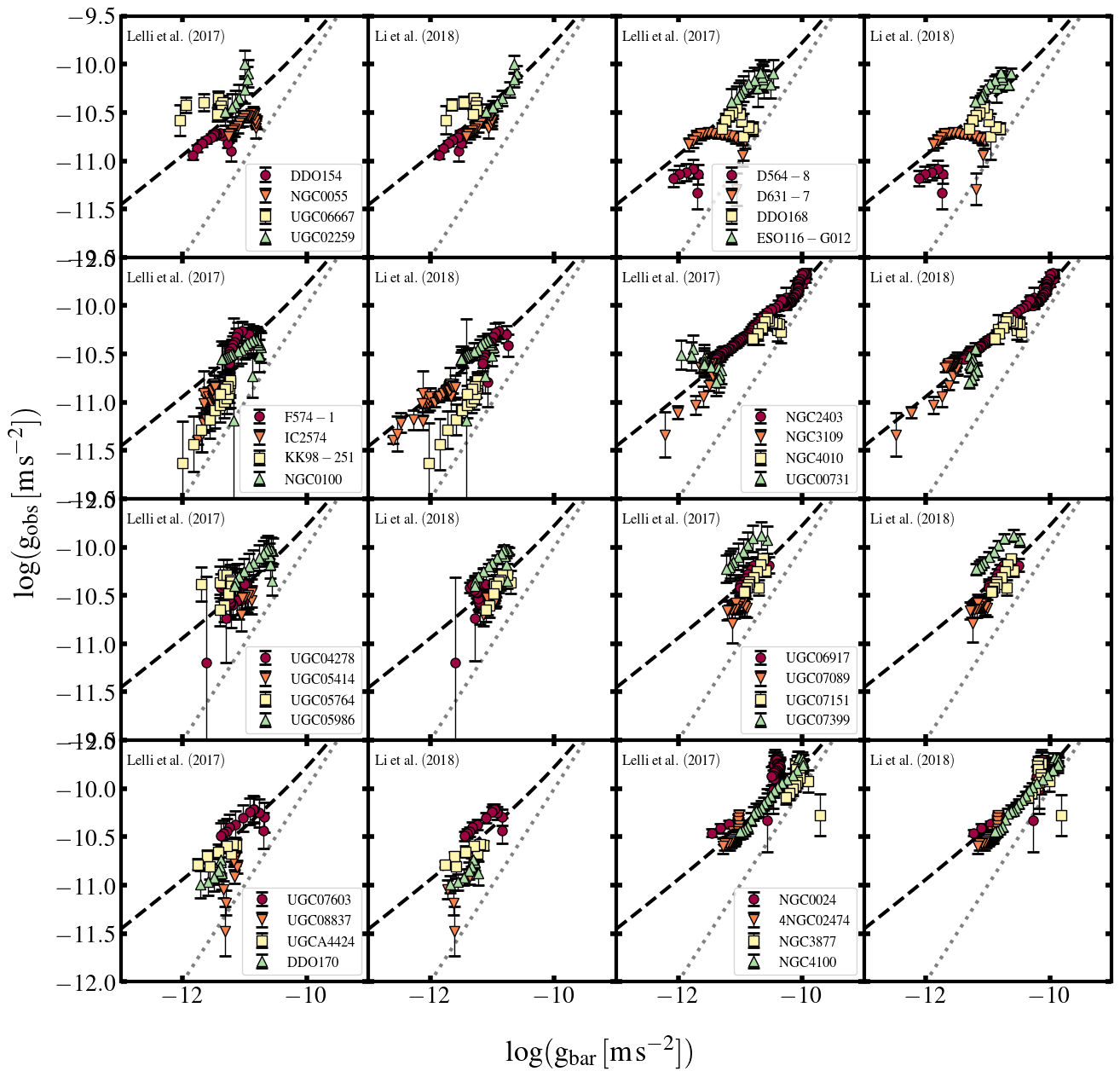}
	\centering
	\caption[]{--- \textit{\textbf{A comparison of individual RAR tracks.}} As in Figure \ref{fig:sparc_hooks}, we provide a side-by-side comparison between individual RAR tracks as presented in \citet{Lelli2017} and the RAR tracks of the same galaxies presented in \citet{Li18} for all 34 galaxies in Table \ref{table:gal_table}. Figure\ref{fig:sparc_hooks} shows the results of the two panels on the top left of this figure. \textit{Takeaway:} For all galaxies except for IC2547 clear non-monotonic behaviour persists even after applying a method that determines galaxy parameters that reduce deviation about the MOND-inspired fit to the observed RAR.}
	\label{fig:AllObsHooks}
\end{figure*}

In Figure \ref{fig:AllObsHooks} we provide a side-by-side comparison between the individual RAR tracks of SPARC galaxies that result from the two methods introduced in \S \ref{sec:ObsHooks}. These comparisons are provided in sets of two panels. For example, the galaxies compared in the two panels of Figure \ref{fig:sparc_hooks} can be seen in the two panels at the top left of Figure \ref{fig:AllObsHooks}. Each pair of panels serves to compare four different galaxies. We find that non-monotonic behaviour persists for all galaxies that we introduce in Table \ref{table:gal_table} even when using values of distance, galaxy inclination and mass-to-light ratio that minimize the deviation about the MOND-inspired fit to the observed RAR.

\section{Total Mass Versus Baryon Mass}
\label{sec:mvm}

\begin{figure*}
	\includegraphics[width=1.0\textwidth, trim = 0 0 0 0]{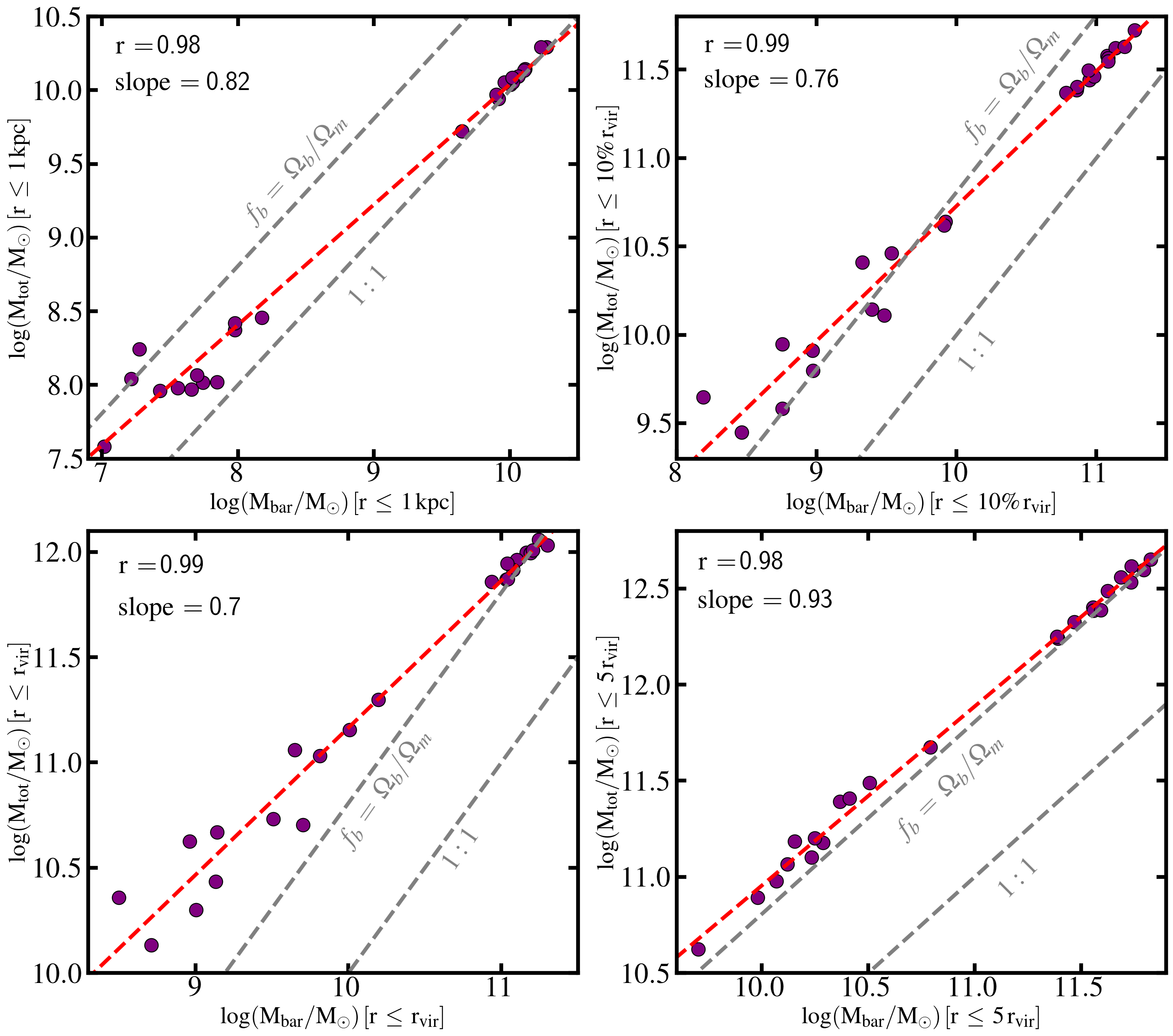}
	\centering
	\caption[]{--- \textit{\textbf{Total versus baryonic mass.}} The total mass as a function of the baryonic mass within 1 kpc (\textit{top left}), 10\% of the virial radius (\textit{top right}), the virial radius (\textit{bottom left}), and five times the virial radius (\textit{bottom right}) for the simulated galaxies in our sample. In each panel two grey dashed lines represent a 1-to-1 relationship (1:1) and another with a normalisation set by the cosmic baryon fraction, $M_{\rm tot} \, \approx \, M_{\rm bar}/\fbar$. The red dashed line represents the least squares fit to the data. Finally, we provide the pearson coefficient as well as the slope of each fitted line at the top left of each panel. \textbf{\textit{Takeaway:}} When measured within the same region, the total and baryonic masses of each galaxy follow very tight power laws with differing slopes depending where the mass is measured. This power law behaviour at different radii dictates the behaviour we see in the RAR.}
	\label{fig:mvm}
\end{figure*}

Here we shine light on the underlying relationship that is responsible for the RAR. In Figure \ref{fig:mvm} we plot the total versus the baryonic mass within 1 kpc ($r \, \leq \, 1 \, \rm kpc$; top left), 10\% of the virial radius ($r \, \leq \, 10\% \, r_{\rm vir} $; top right), the virial radius ($r \, \leq \, r_{\rm vir} $; bottom left), and five times the virial radius ($r \, \leq \, 5 \, r_{\rm vir} $; bottom right) for our simulated galaxies. In each panel, the two grey dashed lines represent a 1-to-1 relationship (1:1) and another with a normalisation set by the cosmic baryon fraction, $M_{\rm tot} \, \approx \, M_{\rm bar}/\fbar$. The red dashed line represents the least squares fit to the data. Finally, we provide the Pearson coefficient as well as the slope of each fitted line at the top left of each panel. When measured within the same region, the total and baryonic masses of each galaxy follow very tight power laws with differing slopes depending where the mass is measured.

Notice that the galaxies follow a power law with a slightly shallower slope than the 1-to-1 relation when their masses are measured within small radii. As we measure the mass within larger radii, the slope of that power law decreases. We posit that this power law slope behaviour at different radii, which is itself a result of a complex combination of competing effects from stellar feedback and gravity, leads to the change in slope of the RAR at accelerations below $a_0$. Finally, when the masses are measured within sufficiently large radii (bottom right), the slope begins to increase until \textit{all} galaxies follow the relation set by the cosmic baryon fraction.

%%%%%%%%%%%%%%%%%%%%%%%%%%%%%%%%%%%%%%%%%%%%%%%%%%

% Don't change these lines
\bsp	% typesetting comment
\label{lastpage}
\end{document}